\documentclass[a4paper,fleqn,usenatbib,useAMS]{mnras}
\usepackage{mathtools, cuted}

\usepackage{graphicx}	
\usepackage{amsmath}	
\usepackage{amssymb}	
\usepackage{bm}		
\usepackage{pdflscape}	

\title{SIToolBox : A package for Bayesian estimation of the isotropy violation in the CMB sky}

\author[S. Das]{Santanu Das$^{1}$$^{2}$\thanks{Contact e-mail: \href{mailto:sdas33@wisc.edu}{sdas33@wisc.edu}}
\\
$^{1}$Department of Physics, University of Wisconsin - Madison, Madison, WI 53706, USA \\
$^{2}$Fermi National Accelerator Laboratory, Batavia, IL 60510, USA
}

\begin{document}
\label{firstpage}
\pagerange{\pageref{firstpage}--\pageref{lastpage}}
\maketitle

\begin{abstract}
The standard model of cosmology predicts a statistically isotropic (SI) CMB sky. However, the SI violation signals are always present in an observed sky-map. 
Different cosmological artifacts, measurement effects and unavoidable effects during data analysis etc. may lead 
to isotropy violation in an otherwise SI sky. Therefore, a proper data analysis technique should calculate all these SI violation signals, so that they can be matched with 
 SI violation signals from the known sources and then conclude if 
there is any intrinsic SI violation in the CMB sky. In one of our recent works, we presented a general Bayesian formalism for measuring the isotropy violation 
signals in the CMB sky in presence of an idealized isotropic noise. In this paper, we have extended the mechanism and develop a software package, 
SIToolBox, for measuring SI violation in presence of anisotropic noise and masking. 

\end{abstract}
\begingroup
\let\clearpage\relax
\endgroup
\newpage

\section{Introduction}
The standard model of cosmology is based on the assumption that CMB sky is statistically isotropic (SI).
However, this assumption of the statistical isotropy has been under intense scrutiny since the plausible detection of 
violation of statistical isotropy in the CMB sky, observed by WMAP mission \citep{Eriksen:2003db}.
Even in the standard model, where the intrinsic CMB sky is SI,
 many external sources 
may introduce SI violation in the observed skymap. SI violation may occur due to weak lensing
\citep{Rotti2011}, Doppler boost due to the motion of our galaxy
with respect to the CMB rest frame 
\citep{Das2015,Hanson2009,Mukherjee2014,Yasini2017,Yasini2016,Yasini2016a} etc. Recent experiments also
show dipole modulation of the low multipoles of the CMB sky due to unexplained 
sources \citep{Ghosh2016a,Fernandez-Cobos2014,Mukherjee2016a}.
Apart from these cosmological effects, there can be many other observational
artifacts. The beam pattern of the CMB experiments are not completely
circularly symmetric due to the unavoidable side lobes etc. The scan
pattern is also not isotropic -- different pixels in the sky get scanned
unevenly from multiple orientations with non-circular beam. 
The noise pattern of the detectors are not 
isotropic. Masking of point sources, galactic plane and other bright regions like LMC, SMC, etc. are necessary during data
analysis. Different foreground removal methods leave some residual foreground signals in
the CMB data. All these introduce the SI violation in the
observed CMB sky \citep{Pant2016,Joshi2012,Das2016,Das2014b,Das2015d,Aluri2015,Aluri2015a}.
Therefore, to detect any intrinsic SI violation, 
we first need to account for all these observational effects. The intrinsic SI violation in the CMB sky may originate 
due to different nonstandard theoretical models, such as
 cosmic topology~\citep{Bond1998,Bond2000}, 
violation of Copernican principle \citep{Ackerman2007,Pullen2007,Lewis2010,Hajian2003a} etc.

Several WMAP
and Planck results recently show SI violation signals of  in the observed maps
\citep{Bennett2011,Bennett2013,Ade2014} and a proper statistical analysis of these data is necessary.
For an isotropic CMB sky, the angular power spectrum is sufficient
to provide the full sky statistics. 
However, the angular power spectrum does not provide any
information about the SI violation. 

In presence of statistical isotropy violation, 
we need the full co-variance matrix
between $a_{lm}$'s (the coefficients of the spherical harmonic
expansion of the sky), i.e. $\left\langle a_{lm} a^*_{l'm'}\right\rangle $, to characterize the full sky statistics.
Here, $\left\langle \,\cdots\,\right\rangle $ denotes the ensemble average of the quantity inside. The co-variance 
matrix, being of the order of $O(l_{max}^2\times l_{max}^2)$, is a really big matrix. Hence, its difficult to infer 
anything directly from the co-variance matrix. 
A better way of quantifying the SI violation in the
CMB sky is to expand the co-variance matrix in terms of the Clebsch-Gordan
coefficients ($C_{lml'm'}^{LM}$) as 
{\small
\begin{equation}
S_{lml'm'} \equiv \left\langle a_{lm} a^*_{l'm'}\right\rangle = 
(-1)^{m'} \sum_{L=0}^{\infty}\sum_{M=-L}^{L}C_{lml'-m'}^{LM}A_{ll'}^{LM}\,.
\label{eq:Eq0}
\end{equation} 
}

\noindent Here, $A_{ll'}^{LM}$'s are known as the BipoSH coefficients.
This method was initially proposed by Hajian and Souradeep \citep{Hajian2003,Hajian2004,Hajian2005},
and is capable of quantifying the SI violation in the CMB sky. For a completely SI sky, apart
from the angular power spectrum, ${\mathcal C}_{l} = (-1)^lA^{00}_{ll}/\sqrt{2l+1}$, all other BipoSH coefficients are $0$.
However, in presence of SI violation, we can detect nonzero signal in other BipoSH coefficients. 

In one of our papers on this topic \citep{Das2015}, we presented a generalized
formalism of estimating the BipoSH coefficients for an idealized sky map with isotropic 
noise, using a completely Bayesian technique.
However, any real CMB data analysis involves anisotropic noise and masking; and 
if it is not properly taken into account,
it may contribute to a false detection of SI violation. 
In some recent works, researchers use bias correction method for calculating the BipoSH coefficients in presence of
anisotropic noise and masking \citep{Aluri2015a,Das2016}.
The method works well and can provide the fairly accurate BipoSH coefficients. 
However, there can be small coupling
between the SI violation from the anisotropic noise and
the SI violation in the intrinsic sky, which can not be accounted
for using a simple bias correction method. Therefore, in this paper we 
develop a software package, known as SIToolBox\footnote{\url{https://github.com/SIToolBox/SIToolBox}}, for jointly calculating the CMB power spectrum and the BipoSH coefficients in presence of
the anisotropic noise and masking. We also recover the Dipole modulation and Doppler boost parameters 
in presence of anisotropic noise. We use different noise pattern and masking to test the efficacy of the algorithm in multiple scenario.   
Our method is completely Bayesian that uses Monte Carlo sampling
for calculating the posterior probability distribution.

The paper is organized as follows. In Sec.~\ref{Sec2}, we describe
the basic mathematics of the BipoSH mechanism and the Bayesian probability distribution 
for the BipoSH coefficients. The third section presents a brief discussion of Hamiltonian Monte Carlo (HMC) method and
describes some of the numerical issues in the data analysis techniques and how to overcome them. In the fourth section we have
given the analysis and results for the BipoSH calculations in presence of anisotropic noise and
masking. In the fifth and sixth section we have calculated the Dipole modulation and Doppler boost terms from anisotropic skymap with masking.
The final section is the discussion and conclusion.

\section{Brief overview of BipoSH formalism} \label{Sec2}

In an ideal CMB observation, our instruments should 
detect the sky temperature at the particular direction of the sky.
However, in reality observe a skymap is a convolution of the instrumental
beam with the sky temperature. Instrumental
noise is also present in the data. The observed sky temperature is given by 

\begin{equation}
d(\gamma_{i})=\int T_i(\gamma_{j})B(\gamma_{i},\gamma_{j})d\Omega_{\gamma_{j}}+n(\gamma_{i}) = T(\gamma_{i})+n(\gamma_{i})\;.
\label{eq:Eq1}
\end{equation}

\noindent Here, $\gamma_{i}$ is the telescope pointing direction. $T_i(\gamma)$, $T(\gamma)$  and $d(\gamma)$
are the real sky temperature, beam-convolved sky temperature and the measured
sky temperature along the direction $\gamma$ respectively. $B(\gamma_{i},\gamma_{j})$
is the beam function and $n(\gamma_{i})$ is the measurement noise 
along $\gamma_{i}$. In a real scan the beam patterns are not symmetric and the 
instrumental noise is not statistically isotropic. Therefore,
these adds isotropy violation features in the observed sky.
As the beam scans any particular direction of the CMB sky multiple times
from different orientations,
it can be considered that each pixel in the sky is getting scanned by an effective beam \citep{Das2016}. This makes it extremely
difficult to de-convolve the beam from the sky map. In this paper we
will not address the problem of de-convolving the beam from the scanned  
skymap. Instead, we will focus on measuring the isotropy 
violation in the beam convoluted skymap, i.e. $ T(\gamma_{i})$. 




For a statistically isotropic skymap the co-variance
matrix is a diagonal matrix and is given by the CMB angular power
spectrum i.e. ${\mathcal C}_{l}$. 
However, in presence of isotropy violation we will
have nonzero values in different BipoSH coefficients as shown in Eq.(\ref{eq:Eq0}). The BipoSH coefficients
not only quantify the SI violation, but do so in a completely
structured manner, i.e. for dipole
modulation, we will see the signal in $A_{ll'}^{1M}$, and
for quadrupolar modulation the signal will be in $A_{ll'}^{2M}$. 

In standard Hajian-Souradeep (HS) format, 
$A_{ll'}^{LM}$s (defined in Eq.(\ref{eq:Eq0})) have an alternating sign for consecutive $l$'s.
Therefore, it is convenient to re-normalize the
BipoSH coefficients as 
\begin{equation}\label{WMAP_norm}
\bar{A}_{ll'}^{LM}=\frac{\sqrt{2L+1}}{\sqrt{2l+1}\sqrt{2l'+1}}\frac{1}{C_{l0l'0}^{L0}}A_{ll'}^{LM}\,.
\end{equation}
This re-normalization was first proposed by the WMAP team and we call
this WMAP re-normalization.   
Under WMAP renormalization $\bar{A}_{ll'}^{LM}$ will have an angular power spectrum like structure making it much easier to interpret. 
In this paper, all the calculations are done in the HS format as they simplifies the calculations. 
However, the results in different figures are presented in the WMAP format as they are easier to interpret visually. 

Given a sky-map, we can expand it in terms of the spherical harmonics and calculate an estimator of the BipoSH coefficients assuming 
that the noise is uncorrelated to the intrinsic sky temperature. 
However, this estimator will be completely dominated by high cosmic variance with signal to noise ratio almost $0$. 
Therefore, it's important to calculate an unbiased estimator of the BipoSH coefficients and calculate the posterior distribution of the estimates. 

The goal of this paper is to sample the the joint probability distribution $P(S_{lml'm'},a_{lm}|d_{lm})$ and get the posterior distribution of the BipoSH coefficients. 
Expanding Eq.(\ref{eq:Eq1})
in spherical harmonics, we get $ d_{lm} = a_{lm} +n_{lm}$, where $n_{lm}$ is the spherical harmonic coefficients of the noise. We can write
\begin{align}
&P(S_{lml'm'},a_{lm}|d_{lm})  =P(d_{lm}|a_{lm})P(a_{lm}|S_{lml'm'})P(S_{lml'm'})\nonumber\\ 
&\propto \frac{1}{\sqrt{|N_{lml'm'}|}}\frac{1}{\sqrt{|S_{lml'm'}|}} \nonumber\\
&\times\exp\left[-\frac{1}{2}\sum_{lml'm'}\left(d_{lm}^{*}-a_{lm}^{*}\right)N_{lml'm'}^{-1}\left(d_{l'm'}-a_{l'm'}\right)\right] \nonumber\\
&\times\exp\left[-\frac{1}{2}\sum_{lml'm'}a_{lm}^{*}S_{lml'm'}^{-1}a_{l'm'}\right]P\left(S_{lml'm'}\right)
\label{lastEquation}
\end{align}

\noindent where $N_{lml'm'}^{-1}$ and $S_{lml'm'}^{-1}$ are the elements
of the inverse of the matrix $N_{lml'm'}$ and $S_{lml'm'}$ respectively
(note that these are not the inverse of the individual elements of the matrix
but the elements of the inverse matrix). $N_{lml'm'}$ is the noise co-variance matrix. 
It's a diagonal matrix for an isotropic noise field. However, for a real scan, 
the noise field being anisotropic the matrix will have all the off diagonal elements. 
For calculating $P(a_{lm}|S_{lml'm'})$ we assume that the $a_{lm}$'s are Gaussian. 
$P\left(S_{lml'm'}\right)$ is the prior on $S_{lml'm'}$. For our analysis we have considered $P\left(S_{lml'm'}\right) = 1$.

\begin{figure*}
\includegraphics[width=0.45\textwidth]{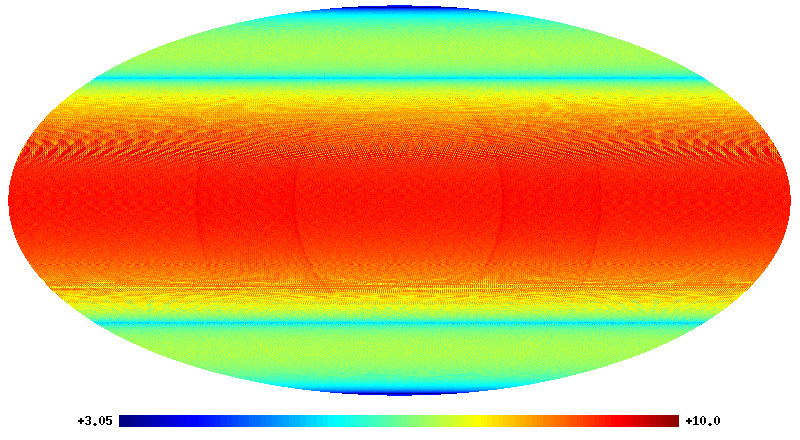}
\includegraphics[width=0.45\textwidth]{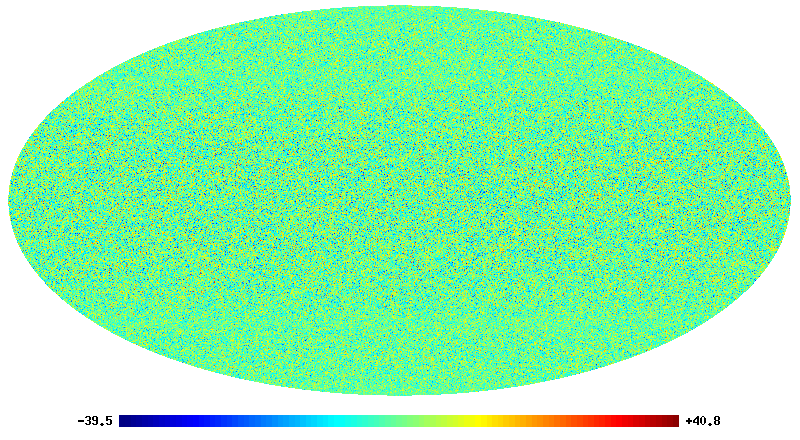}
\caption{Left:The plot shows  $ \sigma_n(\gamma)$, the noise standard deviation map  for $\sigma_n^{max}=10\mu K$.  Right: A sample noise pattern for this particular $\sigma_n(\gamma)$. 
The maps are shown in ecliptic coordinate system.}
\label{fig:NoiseSD}
\end{figure*}
\section{Sampling the distribution}

It's possible to obtain a semi-analytic solution of the probability distribution
of the BipoSH coefficients provided we marginalize over $a_{lm}$s
 and the noise field is isotropic~\citep{Seljak1998}. 
However, without marginalization it's impossible
to get an analytical expression for the full posterior distribution. 
Therefore, we use the Hamiltonian Monte Carlo (HMC) algorithm for drawing samples from $P(S_{lml'm'},a_{lm}|d_{lm})$.

HMC is based on the classical Hamiltonian mechanics and statistical physics. The main idea behind HMC is 
to develop a Hamiltonian function $H( x, p)$ such that the resulting Hamiltonian dynamics allows us to efficiently explore some target distribution $P(x)$.
This can be achieved using a basic concept adopted from statistical mechanics, known as the canonical distribution. The energy function for the Hamiltonian dynamics is a combination of the potential energy $V(x)$ and the kinetic energy $K(p)$ of the system. 
Therefore, the canonical distribution for the Hamiltonian dynamics is  $P(x,p) \propto e^{-H(x,p)} \propto e^{-V(x)}e^{-K(p)}$.  Most important thing here is that  
the joint distribution for $(x, p)$ factorizes. We can use this property to sample any target distribution. 
For sampling a target distribution $P(x)$ we can first define two sets of auxiliary variables, momentum and mass ($p_x$ and $m_x$) corresponding to each $x$. 
The kinetic energy of each micro-state will then be given by $K(p_x) = \frac{p_x^2}{2m_x}$ and we will take $V(x) = -\ln(P(x))$ . This gives  $ e^{-H(x,p)} \propto P(x)e^{-\frac{p_x^2}{2m_x}}$
We can now draw random sample for $p_x$ from a Gaussian distribution with $0$ mean, variance $m_x$ and use Hamiltonian mechanics to evolve the system to a new state.
By repeating the process we can get the full canonical distribution.~\citep{Taylor2008,Hajian2007,DUANE1987216}.  The set $m_x$ is also known as the mass matrix. The final result of the HMC method in independent 
of the choice of the mass matrix. However, a proper choice of $m_x$ is important to stability of the numerical integration for moving from one state $(x, p_x)$ to a new state $(x', p'_x)$~\citep{Das2016}. Theoretically the Hamiltonian should be preserved while going from one state to another. However, due to numerical error the Hamilton do change slightly in the process. Assuming the change of Hamiltonian is $\Delta H$, in HMC the new step is accepted with the probability $\exp(-\Delta H)$. Proper choice of step sizes can make $\Delta H$ significantly small making the acceptance probability $\sim 1$. The step size to achieve acceptance probability $1$ for an $\nu$ th order integrator is analytically calculated in~\citep{Beskos2010}.
We can start HMC with any arbitrary value of $x$. It will first converse close to the best fit value and then it start sampling the probability distribution around it. 
To make the process faster, we can also run multiple chains independently in parallel and combine samples from the chains at the end to obtain the final probability distribution. 


 In our present analysis, 
the parameters are $a_{lm}$ and $A^{LM}_{ll'}$. We define the momentum and mass corresponding 
to these variables as $p_{a_{lm}}$, $m_{a_{lm}}$, $p_{A_{ll'}^{LM}}$ and $m_{A_{ll'}^{LM}}$ respectively. Using these parameters, we can write the Hamiltonian for the HMC sampling as
{\small
\begin{equation}
H = \sum_{lm}\frac{p^2_{a_{lm}}}{2m_{a_{lm}}}+\sum_{LMll'}\frac{p^2_{A^{LM}_{ll'}}}{2m_{A^{LM}_{ll'}}} - \ln(P(S_{lml'm'},a_{lm}|d_{lm}))\;.
\end{equation}
}

\noindent Using classical Hamiltonian mechanics, the equation of motion for the HMC
sampling can be obtained 
as~\citep{Das2015} 

{\small
\begin{equation}
\dot{p}_{a_{lm}}=-\sum_{l'm'}S_{lml'm'}^{-1}a_{l'm'}^{*}+\sum_{l'm'}N_{lml'm'}^{-1}\left(d_{l'm'}^{*}-a_{l'm'}^{*}\right)\;,\label{eq:p_dot_a_lm}
\end{equation}
}
{\small
\begin{equation}
\dot{p}_{A_{ll'}^{LM}}=-\frac{1}{2}\partial_{A_{ll'}^{LM}}\ln\left|S\right|+\partial_{A_{ll'}^{LM}}\left(\sum_{lml'm'}a_{lm}^{*}S_{lml'm'}^{-1}a_{l'm'}\right)\label{eq:Eq8}
\end{equation}
}
\noindent and
 
\begin{equation}
\dot{a}_{lm} = {p}_{a_{lm}}/{m}_{a_{lm}}\;,\label{eq:dot_a_lm}
\end{equation}
\begin{equation}
\dot{A}_{ll'}^{LM} = {p}_{A_{ll'}^{LM}}/{m}_{A_{ll'}^{LM}}\;.\label{eq:dot_A_lm}
\end{equation}

\noindent The partial derivatives with respect to $A_{ll'}^{LM}$ can be calculated as

\begin{align}
&\partial_{A_{ll'}^{LM}}\left( \sum_{lml'm'}a_{lm}^{*}S_{lml'm'}^{-1}a_{l'm'}\right) \nonumber \\ 
&\;\;\;\;\;\;\;\;\;\;=\sum_{mm'}C_{lml'm'}^{LM}\left(S^{-1}a\right)_{lm}\left(S^{-1}a\right)_{l'm'}\;,\label{eq:Eq9}
\end{align}

\begin{equation}
\partial_{A_{ll'}^{LM}}\ln\left|S\right|=\sum_{mm'}C_{lml'm'}^{LM}S_{lml'm'}^{-1}\;.
\end{equation}

HMC is performed in two steps. First, the values of the momentum variables are chosen from the Gaussian distribution of mean $0$ and variance $m_x$, where $x \in (a_{lm}, A^{LM}_{ll0})$. Next, we integrate the equations of motion through a time interval of $\Delta t$, to go from the state ($p_{a_{lm}}$, $p_{A^{LM}_{ll'}}$, $a_{lm}$, $A^{LM}_{ll'}$) to a new state
($p^{*}_{a_{lm}}$, $p^{*}_{A^{LM}_{ll'}}$, $a^{*}_{lm}$, $A^{*LM}_{ll'}$). This concludes one HMC step. Then, we choose another set of Gaussian random momentum and repeat the process. The values of $a_{lm}$ and $A^{LM}_{ll'}$ are stored after each integration step and the samples will follow the posterior distribution of the respective variables. The integration time length $\Delta t$ is varied between steps to avoid resonance. 

\begin{figure*}
\includegraphics[width=0.48\textwidth,trim = 0 50 0 50, clip]{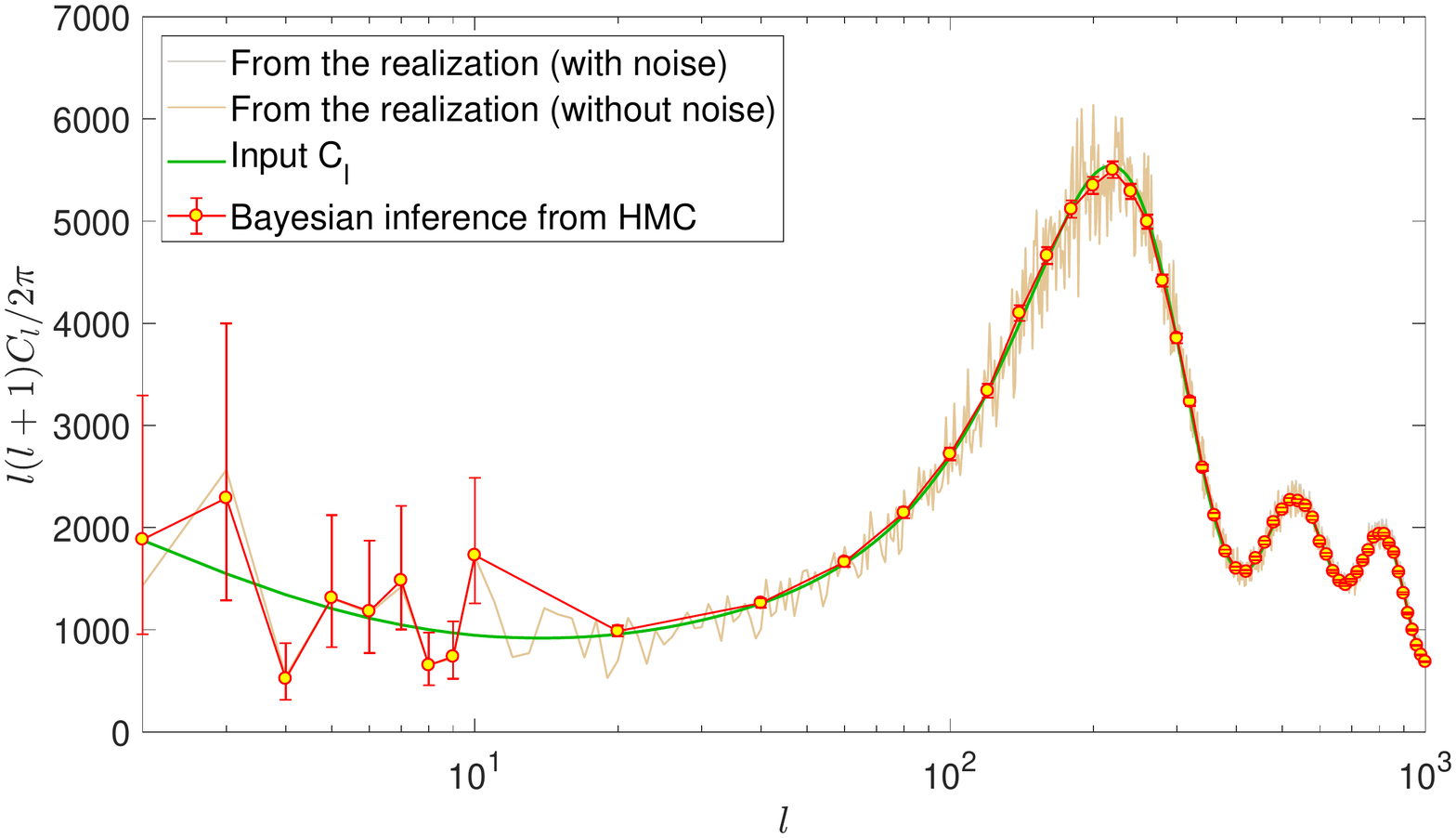}
\includegraphics[width=0.48\textwidth,trim = 0 50 0 50, clip]{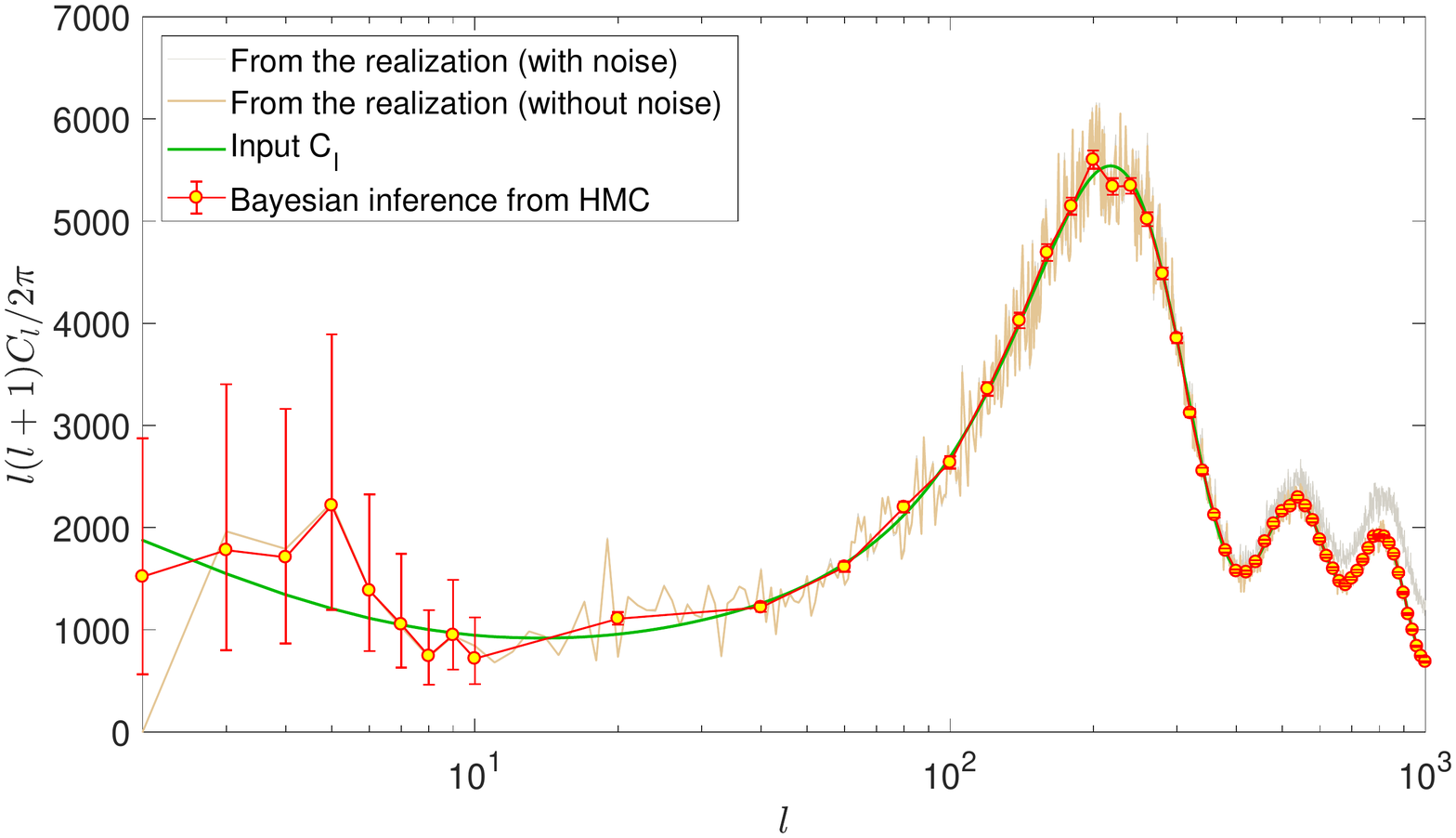}
\includegraphics[width=0.48\textwidth,trim = 0 50 0 50, clip]{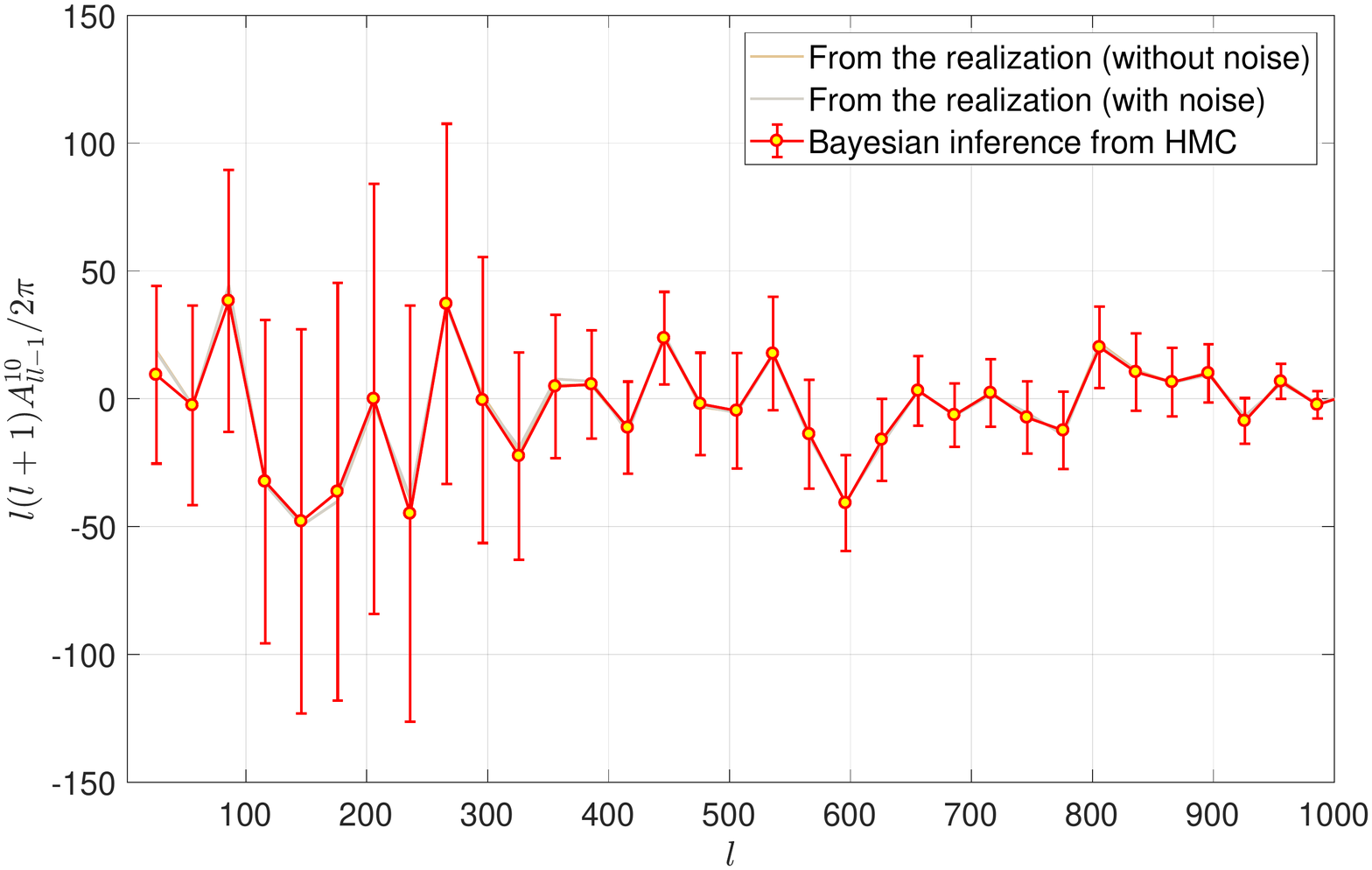}
\includegraphics[width=0.48\textwidth,trim = 0 50 0 50, clip]{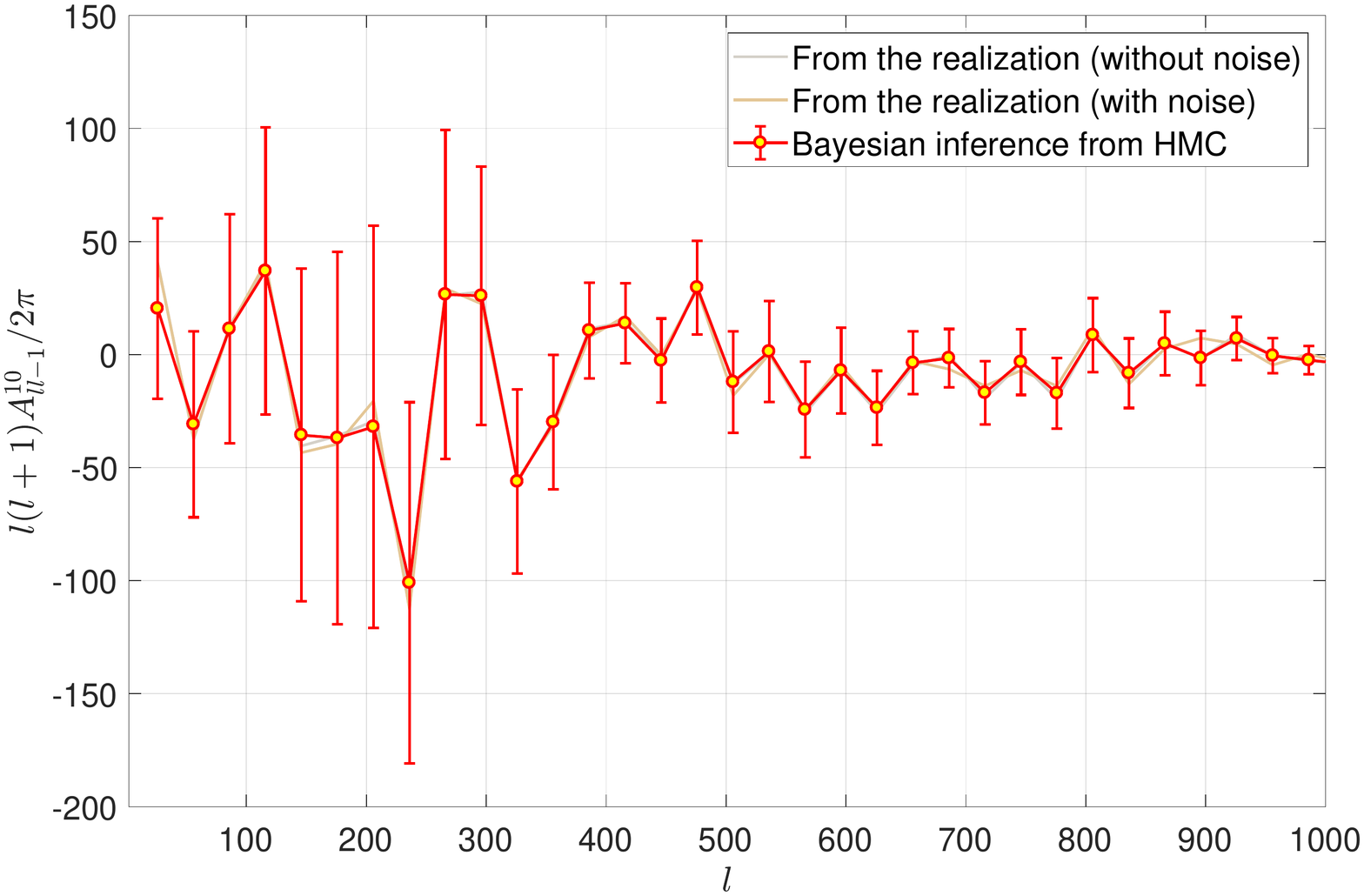}
\includegraphics[width=0.48\textwidth,trim = 0 50 0 50, clip]{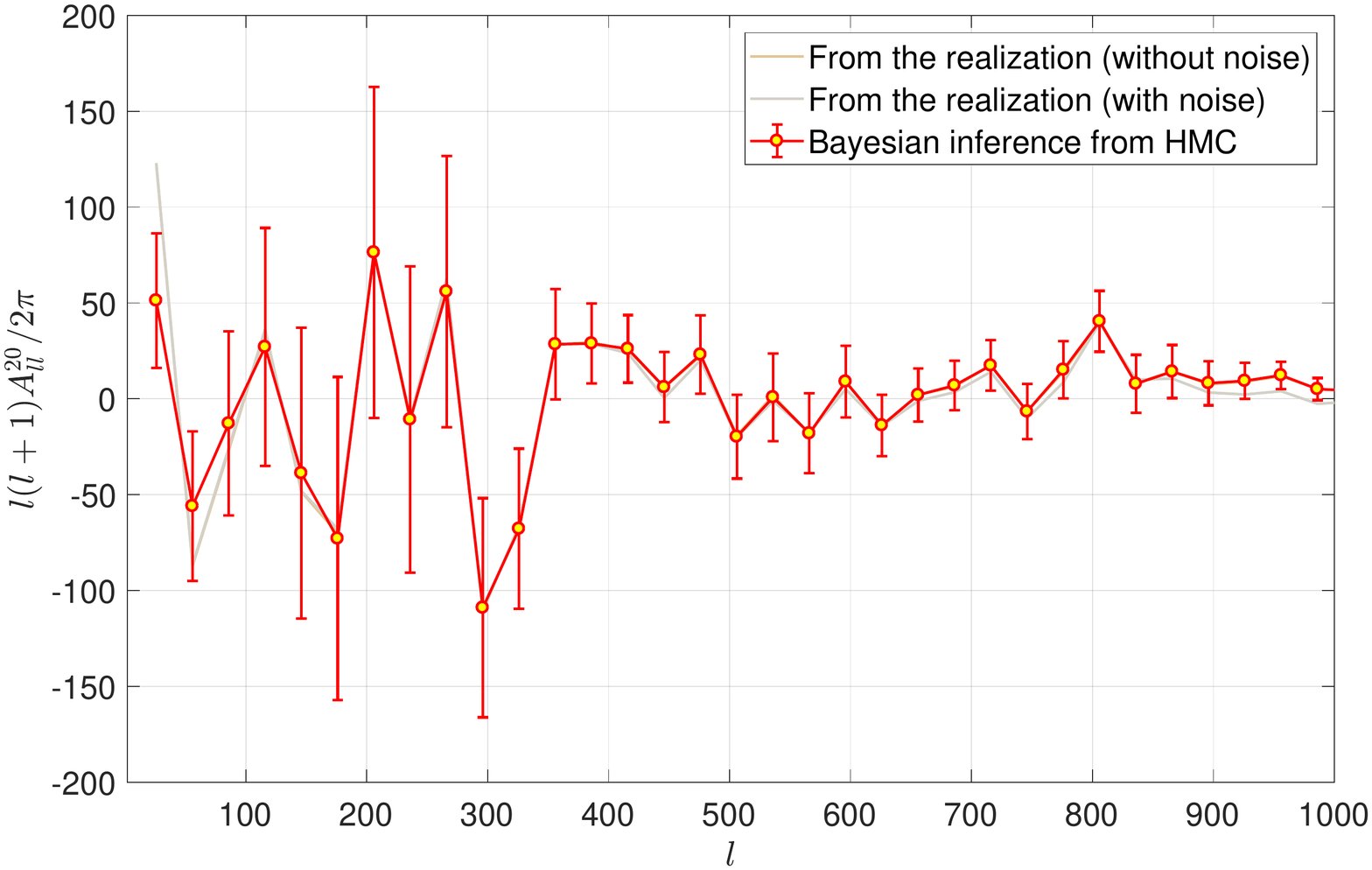}
\includegraphics[width=0.48\textwidth,trim = 0 50 0 50, clip]{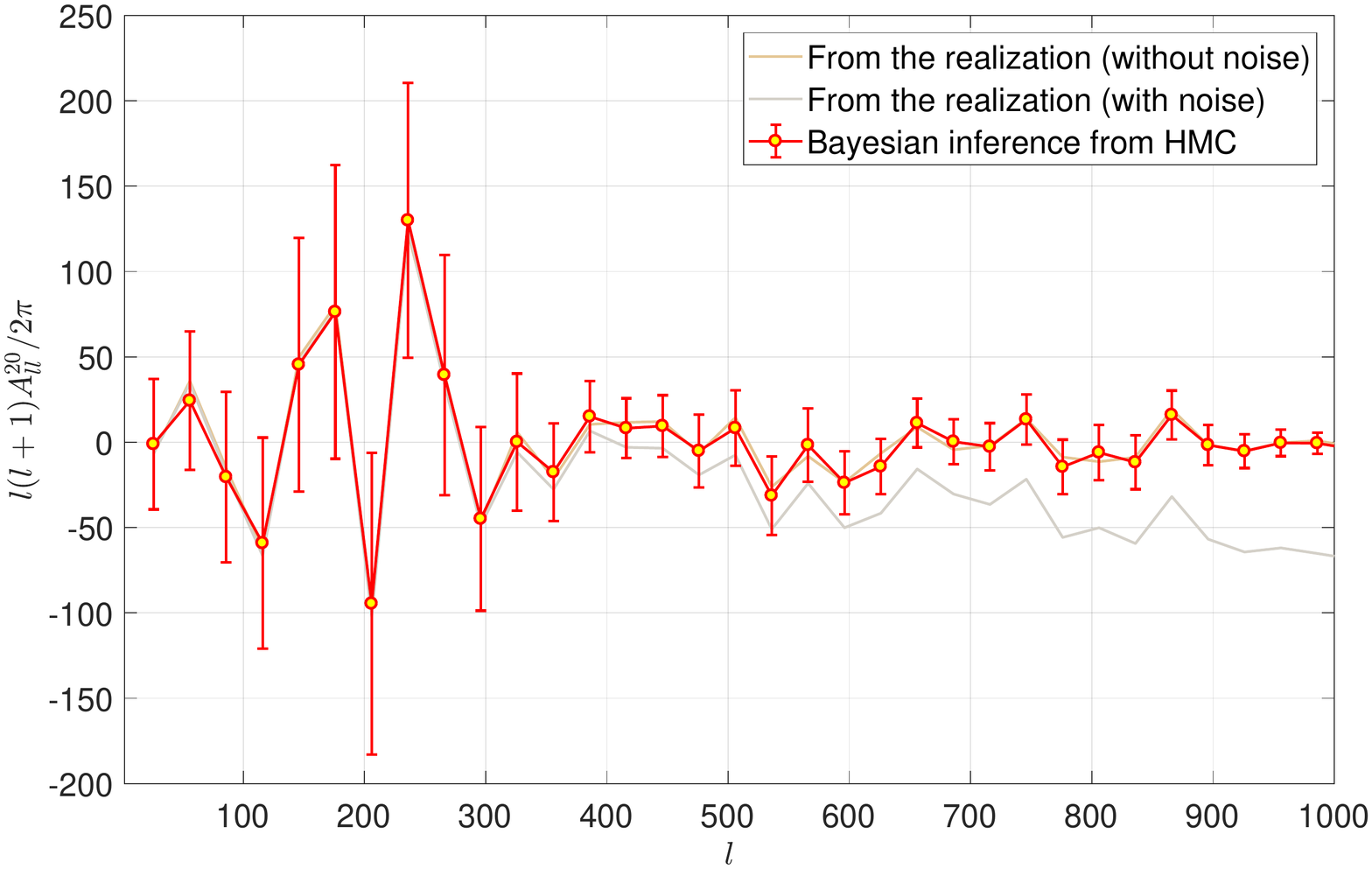}
\includegraphics[width=0.48\textwidth,trim = 0 50 0 50, clip]{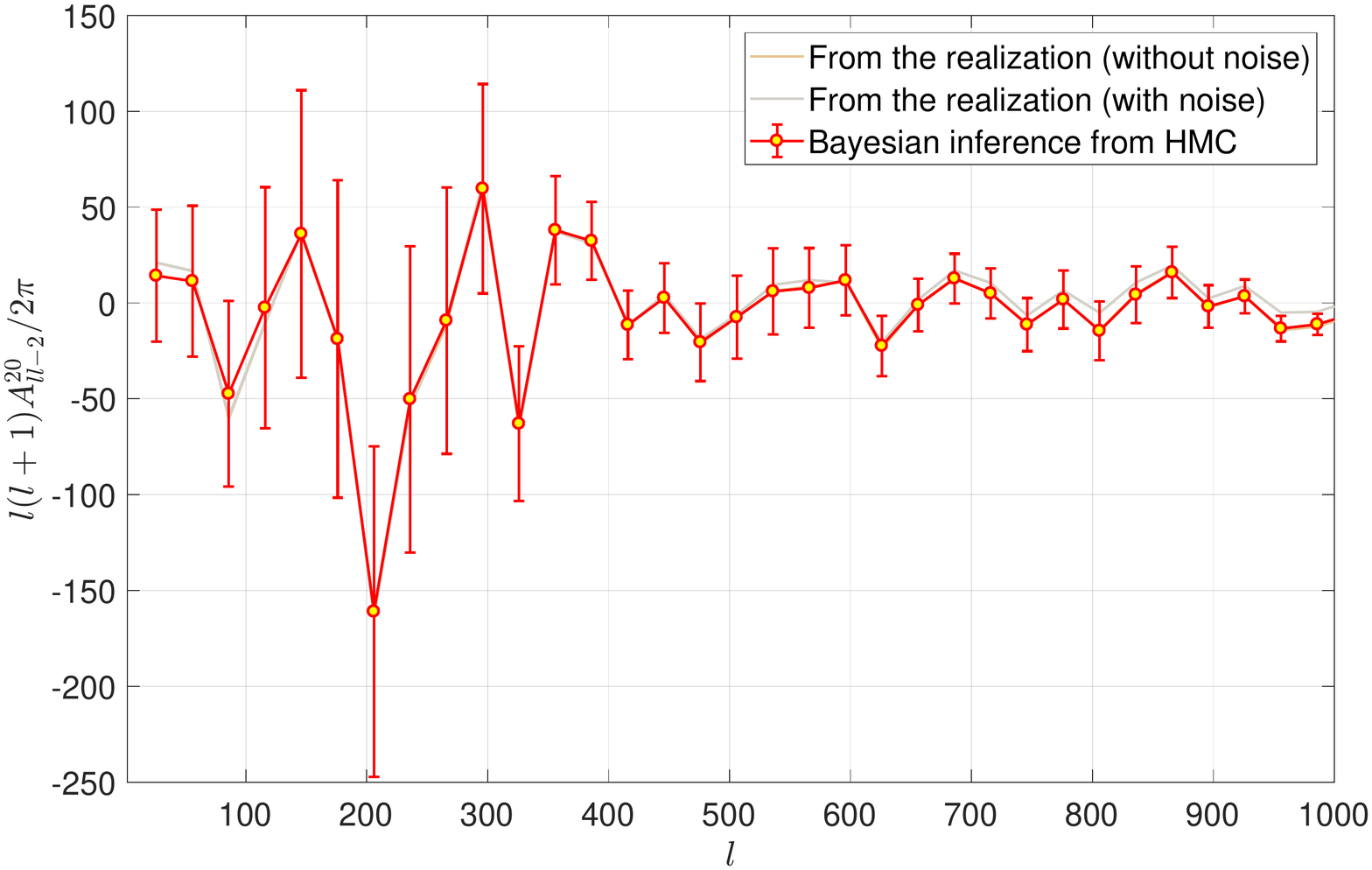}
\includegraphics[width=0.48\textwidth,trim = 0 50 0 50, clip]{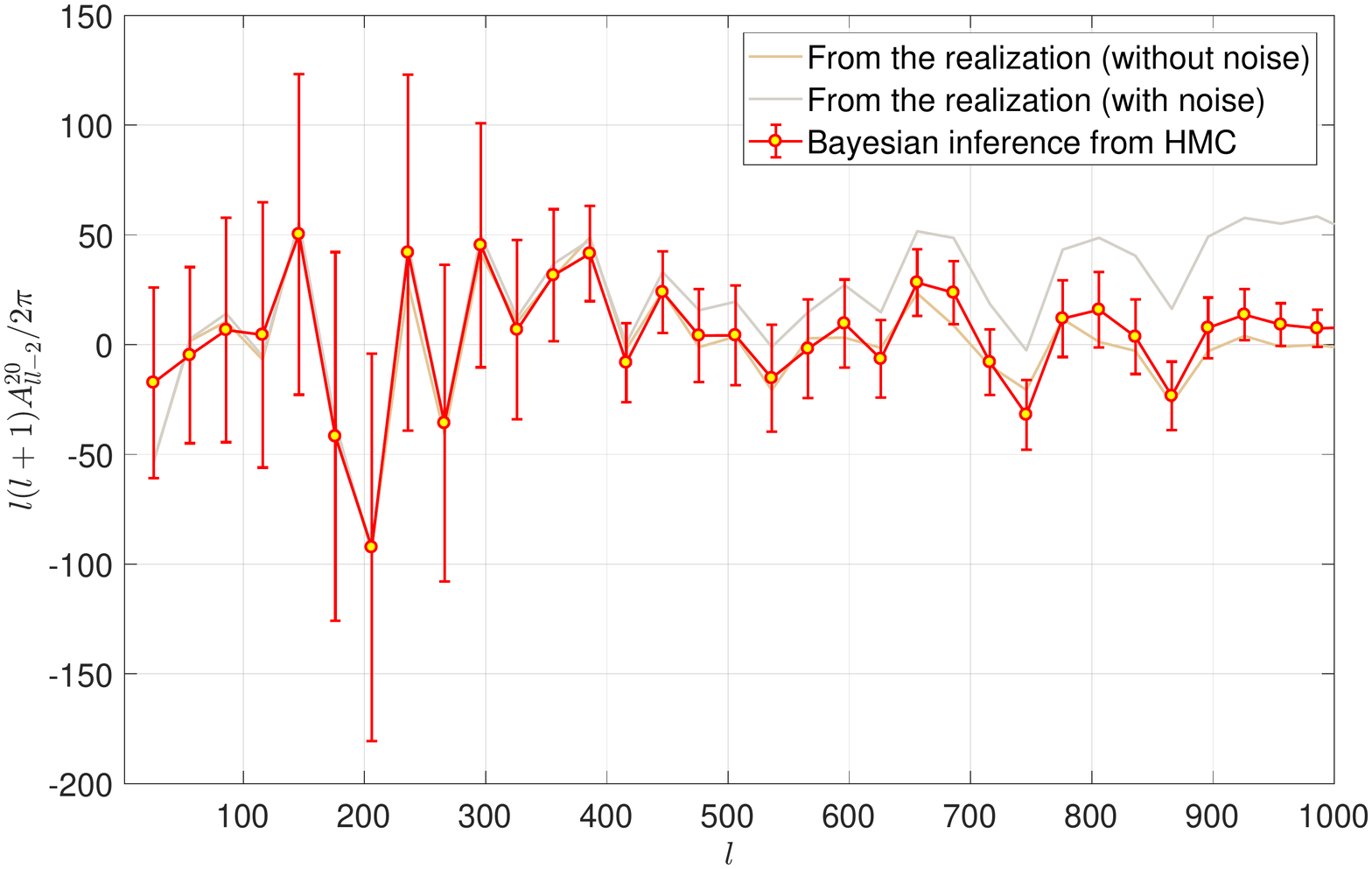}
\caption{Analysis with SI CMB map and anisotropic noise. 
Left : Plots for $\sigma_n^{max} = 10\mu K$, Right: Plots for $\sigma_n^{max} = 30\mu K$. 
Top: Plots of the angular power spectrum. Next plots are for $A^{10}_{ll-1}$,  $A^{20}_{ll}$, $A^{20}_{ll-2}$ respectively. 
Red curve with errorbars are the recovered values from Bayesian inference. Gray and Ochre curves represent values from the input realizations 
before and after adding noise. All the BipoSH coefficients shown in this figure are WMAP normalized.}
\label{fig:isotropicmap}
\end{figure*}
\subsection{Computing the inverse of the noise matrix}

The noise matrix, i.e. $N_{lml'm'}$ is a large matrix, of the order of O$(l_{max}^4)$ and depending on the scan-pattern 
and instrumental noise, the off-diagonal terms of  $N_{lml'm'}$ 
can be large in comparison to the diagonal terms. Therefore,
it's not straight forward to invert the matrix. As the matrix is not  
diagonally dominated, a Taylor series expansion for the $N_{lml'm'}$ around some diagonal matrix is not possible.
The size of the matrix being of the order of $10^6 \times 10^6$ for $l_{max} = 1024$, it's also impossible to store the full matrix. 
However, under the assumption of white noise, the noise co-variance matrix 
is a diagonal matrix in the pixel space. Hence, inverting
the co-variance matrix in pixel space is same as inverting the individual elements of the matrix. 
Even in case of weakly correlated noise, 
the pixel space noise co-variance matrix is diagonal dominated. Hence it can be expanded into
Taylor series and inverted without much computation. 
 
In our calculations,
we convert the map from $a_{lm}$ space to pixel space
and then multiply it with the inverse of the noise co-variance matrix, and  convert it back to the spherical
harmonic space. 
This reduces the computational cost of inverting the $N_{lml'm'}$
matrix. 

\subsection{Computing the inverse of the $S_{lml'm'}$ matrix}

Another challenging task is to invert the $S_{lml'm'}$ matrix, which is not diagonal.
Assuming that the CMB sky is mostly isotropic, $S_{lml'm'}$ matrix is diagonal
dominated. 
Therefore, any numerical inversion technique
like Gauss Seidel method works perfectly for this inversion~\citep{Das2015}.
However, a Taylor series expansion also works very well for this case.
Even if we expand the terms up to the first order terms, we can get
a very good approximation of the results. 

For the Taylor series expansion we write $\left(S_{lml'm'}\right)=D_{lml'm'}+O_{lml'm'}$.
Here $D_{lml'm'}$ is a diagonal matrix only consists of $C_{l}$
and rest is taken as $O_{lml'm'}$. All the terms of $O_{lml'm'}$
are significantly smaller in comparison with $D_{lml'm'}$. Therefore,
expanding $S_{lml'm'}$ in terms of the Taylor series, we get $\left(S_{lml'm'}\right)^{-1}=\left(D_{lml'm'}\right)^{-1}-\left(D_{lml'm'}\right)^{-1}\left(O_{lml'm'}\right)\left(D_{lml'm'}\right)^{-1}$

Few algebraic manipulation gives us
\begin{eqnarray}
\partial_{A_{ll'}^{LM}}\ln\left|S\right|&=&\left(-1\right)^{L+l+l'+1}\sqrt{\left(2l+1\right)\left(2l'+1\right)} \nonumber \\
            &\times&A_{ll'}^{LM}/\left(A_{ll}^{00}A_{l'l'}^{00}\right)\;.
\end{eqnarray}

\noindent Similarly expanding Eq.(\ref{eq:Eq9}) up to first
order, we get

{\small
\begin{align}
&\sum_{mm'}C_{lml'm'}^{LM}\left(S^{-1}a\right)_{lm}\left(S^{-1}a\right)_{l'm'} = \left(-1\right)^{L+l+l'+1} \nonumber\\
&\;\;\;\times\frac{\sqrt{\left(2l+1\right)\left(2l'+1\right)}}{\left(A_{ll}^{00}A_{l'l'}^{00}\right)} \left(\sum_{mm'}C_{lml'm'}^{LM}a_{lm}a_{l'm'}\right)\;.
\end{align}
}

\noindent 
Its possible of expand both the equations up to second order or higher without much complication. However, 
our test results show that the first order approximations work well for the $S$ matrix inversion. 

\begin{figure*}
\includegraphics[width=0.49\textwidth,trim = 0 50 0 50, clip]{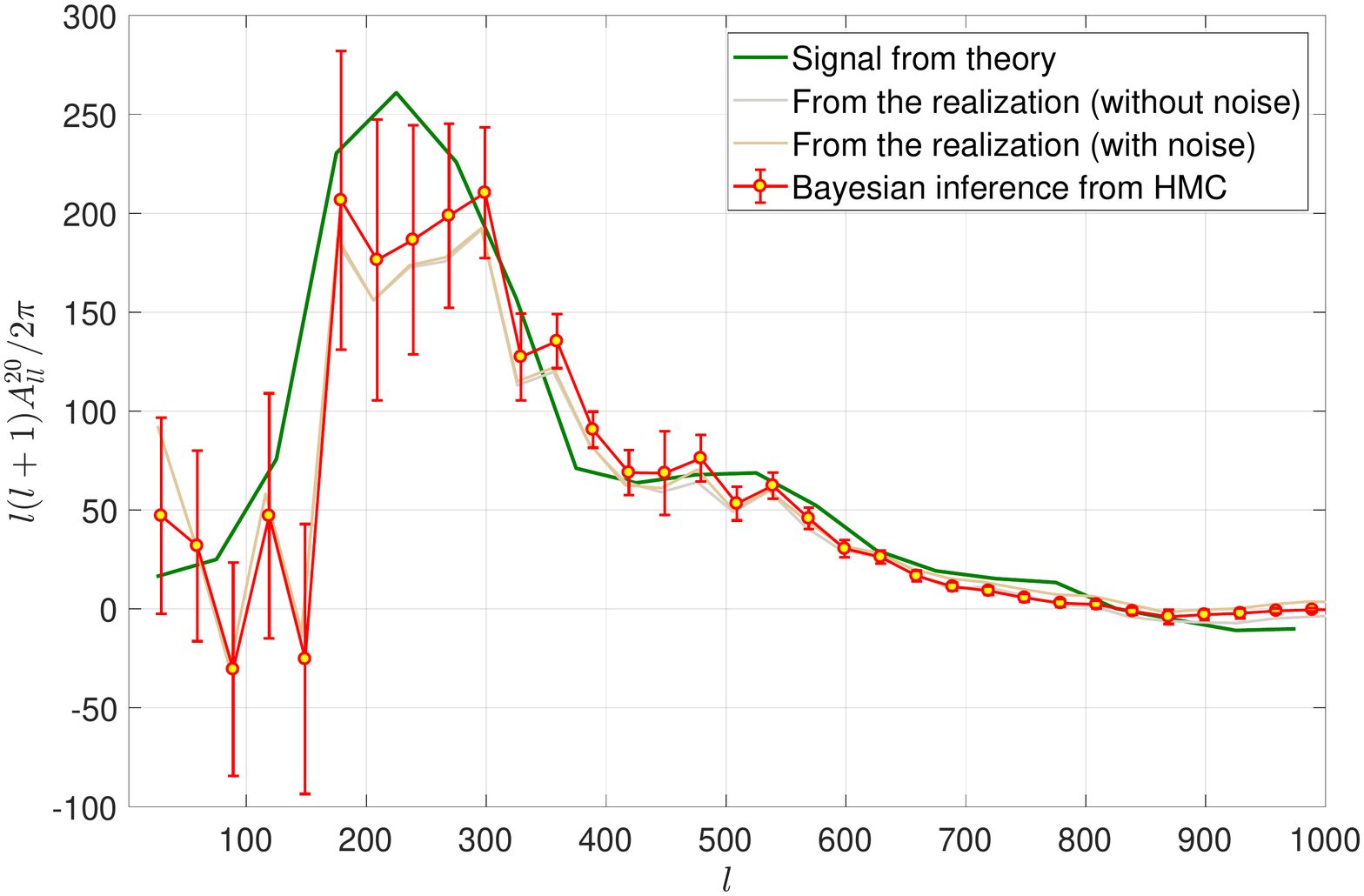}
\includegraphics[width=0.49\textwidth,trim = 0 50 0 50, clip]{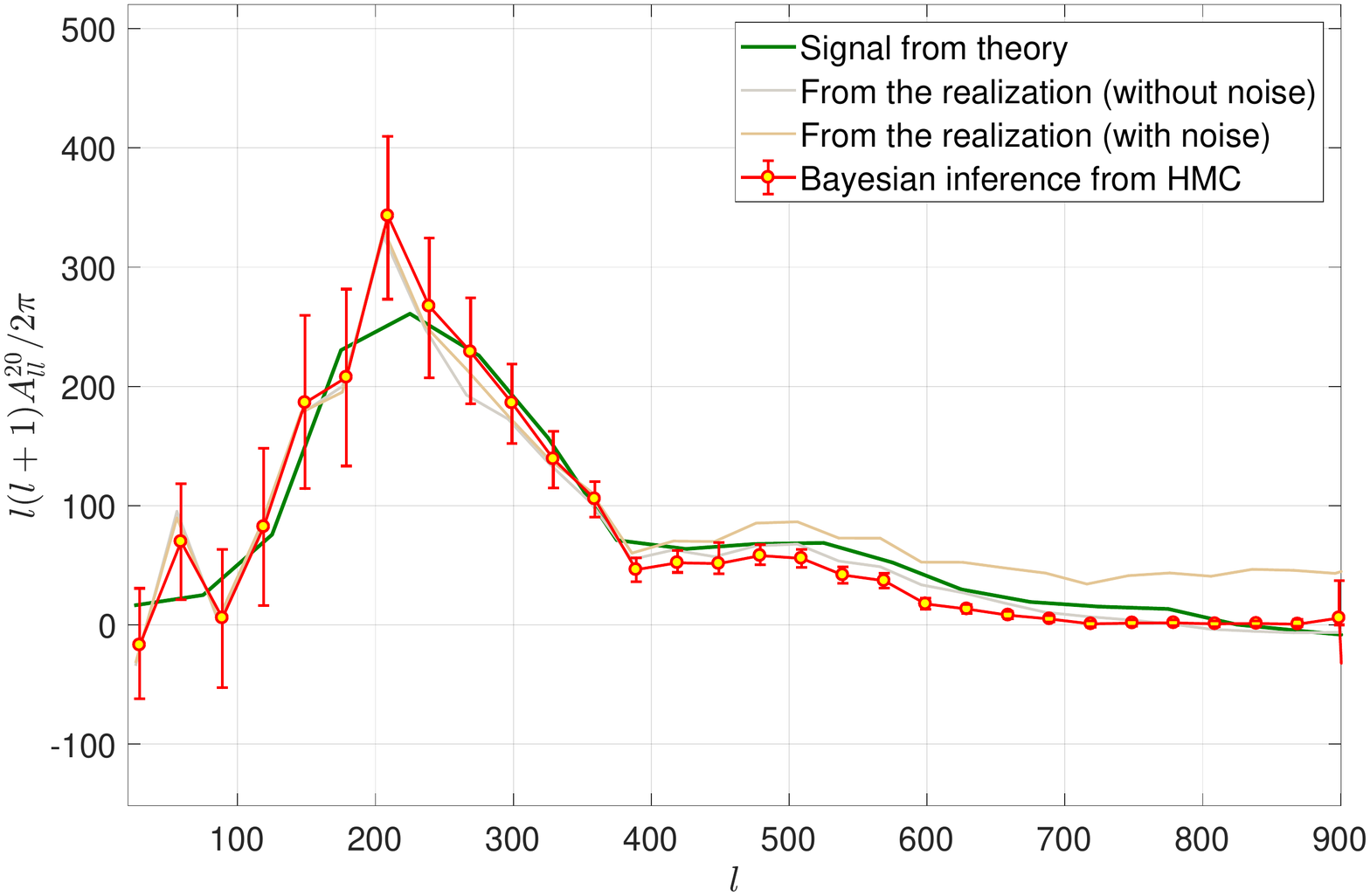}
\includegraphics[width=0.49\textwidth,trim = 0 50 0 50, clip]{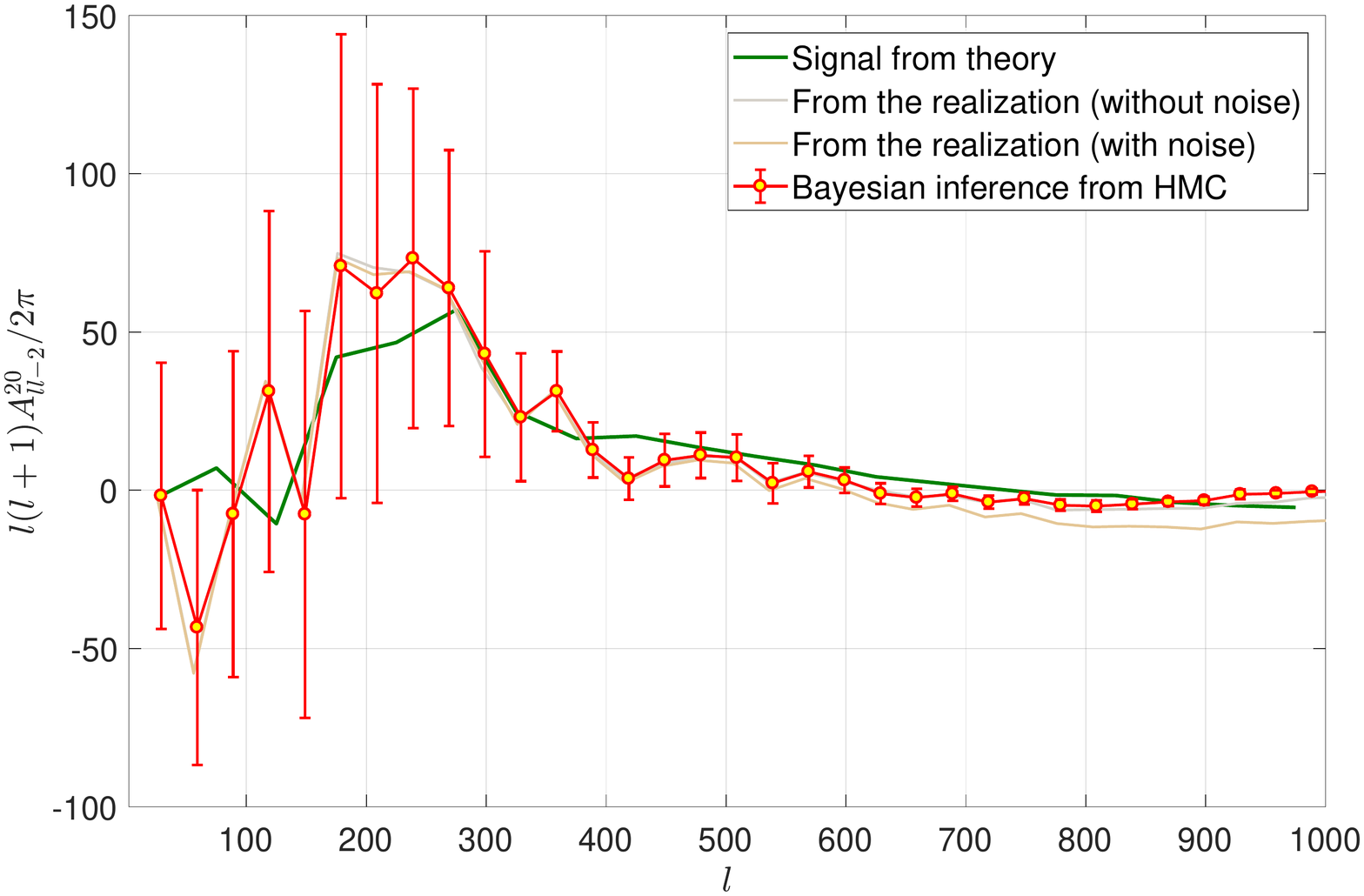}
\includegraphics[width=0.49\textwidth,trim = 0 50 0 50, clip]{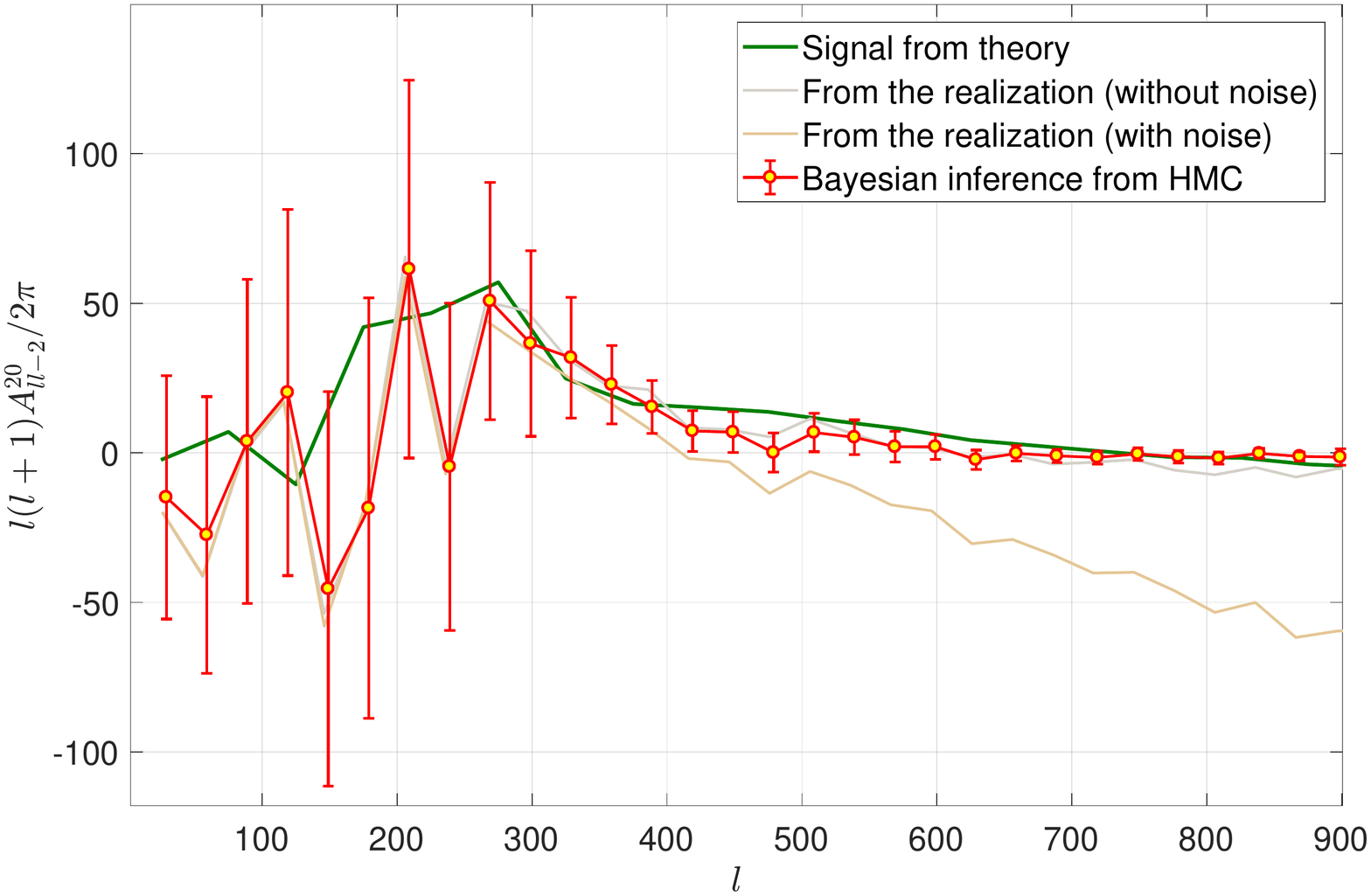}
\caption{Analysis with nSI CMB map generated by scanning a SI realization  with WMAP W2 beam and scan pattern and an anisotropic noise field added to it. 
Left : Plots for $\sigma_n^{max} = 10\mu K$, Right: Plots for $\sigma_n^{max} = 30\mu K$. 
Top and the bottom rows are for  $A^{20}_{ll}$, $A^{20}_{ll-2}$ respectively.  The BipoSH coefficients shown in this figure are WMAP normalized.}
\label{fig:anisotropicmap}
\end{figure*}

\subsection{Stability of the algorithm and the mass matrix}
One of the most challenging problem in our algorithm is the stability of the integration process.  Even though the Leapfrog integrator is common in Hamiltonian Monte Carlo algorithm due to its symplectic nature, the propagation error is large. This will change the value of the Hamiltonian ($\Delta H$) significantly, from one step to another provided we use large time step in the integration process. Therefore, we use a
fourth order symplectic integrator, namely Forest and Ruth integrator that allows us to choose larger step size in the numerical integration process. Our analysis shows that this particular integrator performs much better than the standard Leapfrog method. 

A proper choice of the mass matrix is also crucial for the stability of the numerical integration of Eq.(\ref{eq:p_dot_a_lm}) - Eq.(\ref{eq:dot_A_lm}) in HMC. Our numerical stability analysis, presented in \citep{Das2015}, shows that a mass matrix $m_{a_{lm}}=(C_l^{-1}+N_l^{-1})^{-1}$ ensures the stability of the integration process. 
However, this particular choice of mass matrix with an anisotropic noise field, will 
provide a non-diagonal mass matrix which will add an extra complexity to the problem.  

Therefore, in our present analysis we use same mass matrix as the isotropic noise case, where in stead of $N_l$, we use $N^{(max)}_l$. $N^{(max)}_l$ is the noise variance for a isotropic noise field, where noise variance in pixel space equal to the maximum noise variance of the anisotropic noise field in pixel space. We test that choice of this particular mass matrix does not affect the integration accuracy significantly. 

Mass matrix for $A^{LM}_{ll'}$ is taken as the inverse of their theoretical variance, i.e. $M_{A^{LM}_{ll'}} = \Bigg|\frac{\sqrt{(2l+1)(2l'+1)}}{2A^{LM}_{ll}A^{LM}_{l'l'}}\Bigg|$ and it works well for anisotropic skymap.

\section{Demonstration of the method on simulated CMB sky}

\subsection{SI skymap and Anisotropic noise field}
\begin{figure*}
\includegraphics[width=0.49\textwidth,trim = 0 0 0 0, clip]{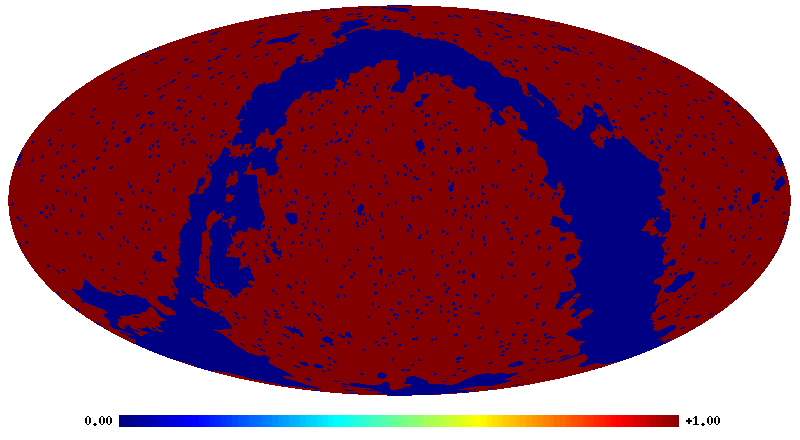}
\includegraphics[width=0.49\textwidth,trim = 0 0 0 0, clip]{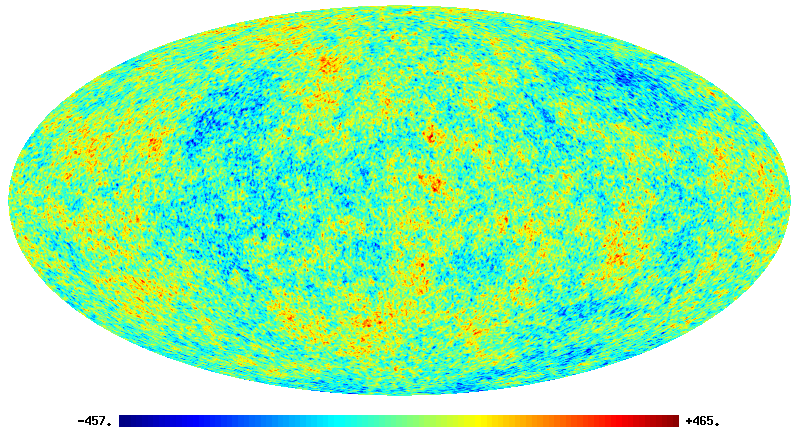}
\includegraphics[width=0.50\textwidth,trim = 0 0 0 0, clip]{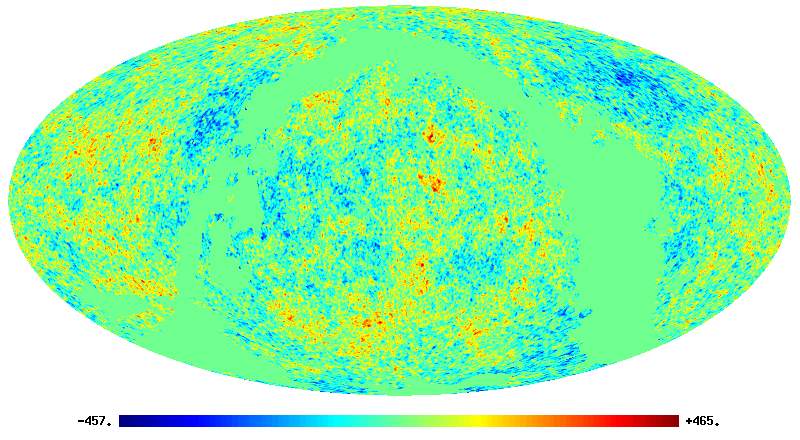}
\includegraphics[width=0.48\textwidth,trim = 0 0 0 0, clip]{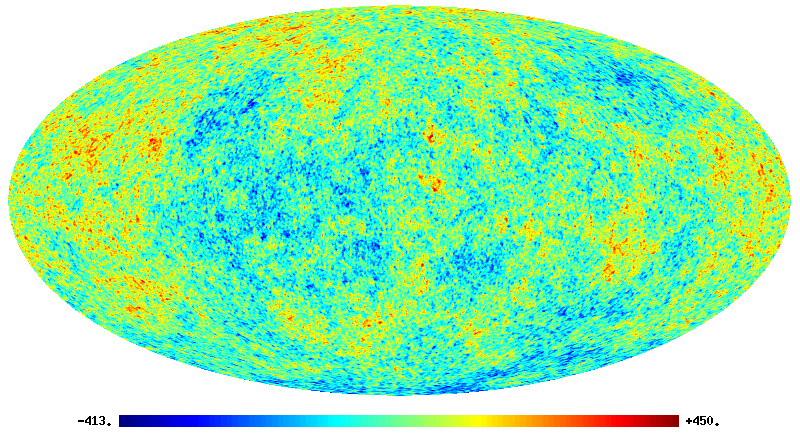}

\caption{Analysis with a masked nSI CMB map generated by scanning a SI realization  with WMAP W2 beam and scan pattern and an anisotropic noise field added to it. 
Top Left : Mask map, used from the analysis. Top Right: Input skymap generated from the time ordered data. Bottom Left: Skymap after adding noise and masking. The shape of the noise standard deviation is same as shown in Fig.~\ref{fig:NoiseSD} ($\sigma_n^{max} = 30\mu K$). Bottom Right: One of the realization recovered from our analysis.}
\label{fig:anisotropicmaskmap_maps}
\end{figure*}

\begin{figure*}
\includegraphics[width=0.49\textwidth,trim = 0 300 0 300, clip]{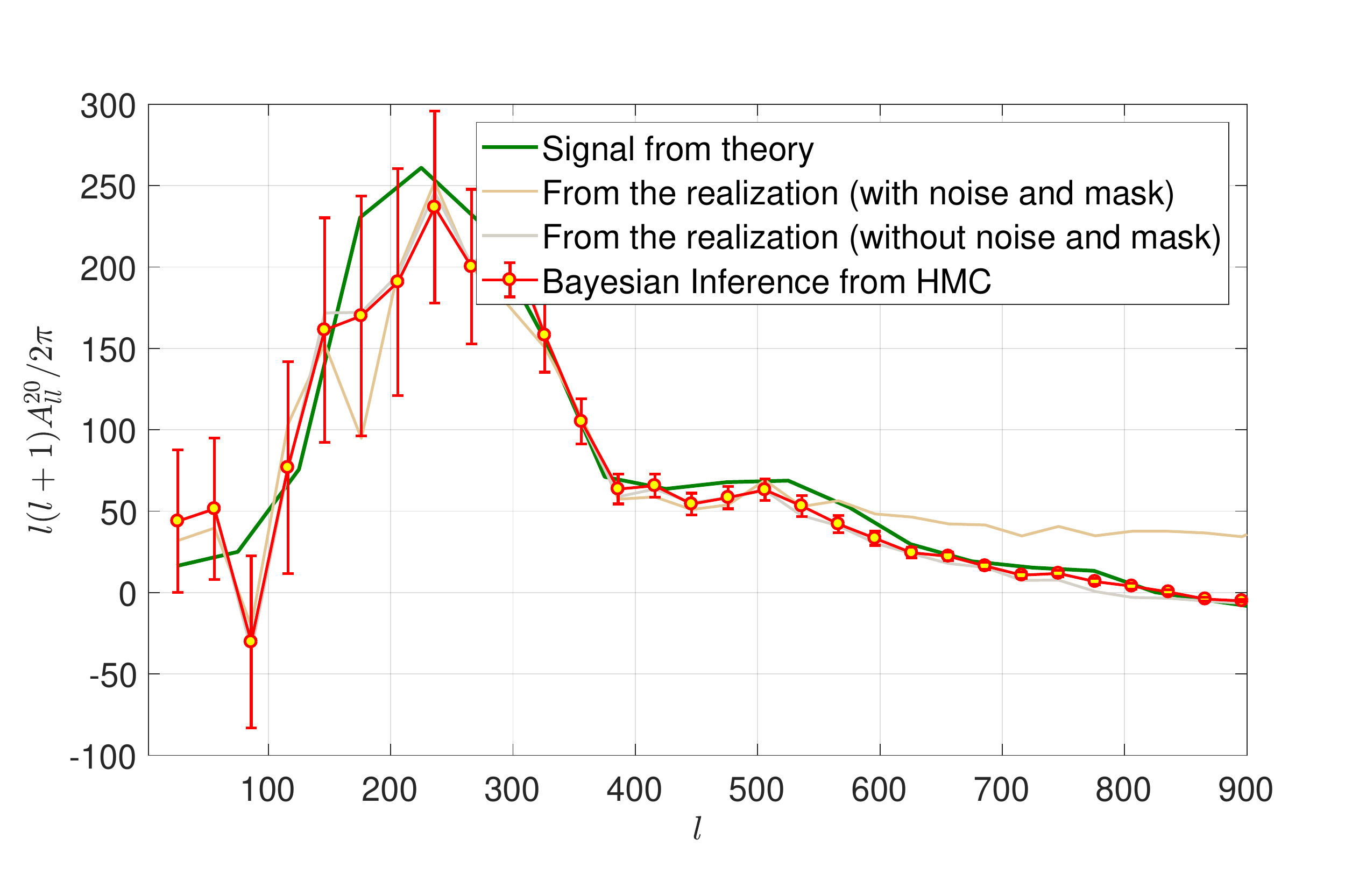}
\includegraphics[width=0.49\textwidth,trim = 0 300 0 300, clip]{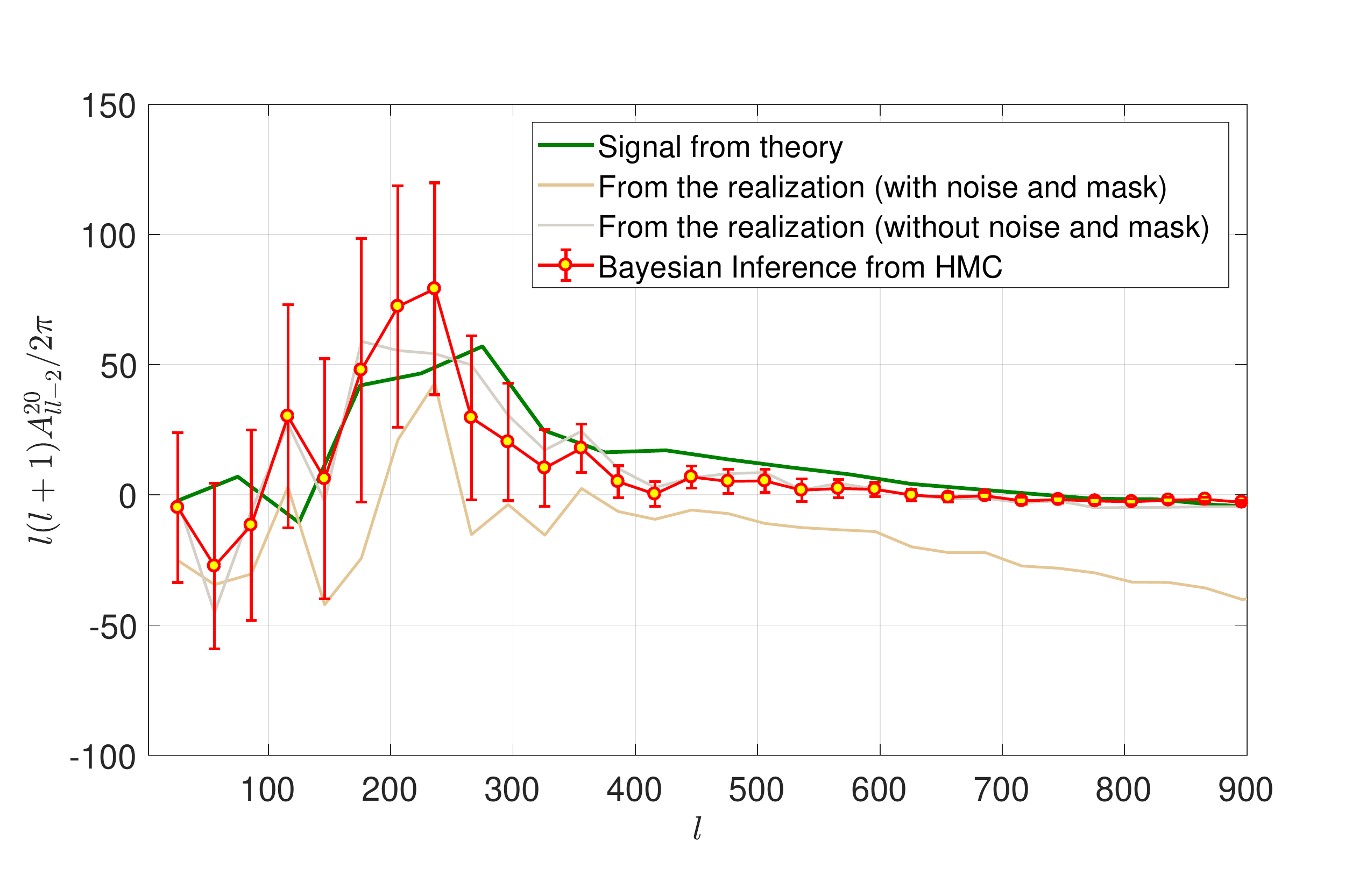}
\caption{Analysis with nSI CMB map in presence of anisotropic noise and masking ($ \sigma_n^{max} = 30\mu K$). 
We show the plots for  $A^{20}_{ll}$, $A^{20}_{ll-2}$.  The BipoSH coefficients shown in this figure are WMAP normalized.}
\label{fig:anisotropicmaskmapspectra}
\end{figure*}

 In any satellite based experiments like WMAP or Planck, all the pixels in the sky don't get scanned equal number of times. We assume that the noise standard deviation at any pixel, $\sigma_n(\gamma)$\footnote{$\sigma_n(\gamma)$ is used for the noise standard deviation in the pixel space and the pixel space noise variance is represented by $N(\gamma) = \sigma_n^{2}(\gamma)$.}, is inversely proportional to the 
square root of number of hits. Mathematically saying, 
$\sigma_n(\gamma) = \sigma^{max}_n \times {\mathcal F}(\gamma)/{\mathcal F}_{max}(\gamma)$ where ${\mathcal F}(\gamma)=1/\sqrt{H(\gamma)}$. $H(\gamma)$ is the number of times a pixel along $\gamma$ direction gets scanned (hit count). 

We generate a SI skymap using HEALPix~\citep{Gorski2004} with $N_{side}=512$. No beam is considered for this analysis.
We take a WMAP like scan pattern for generating the noise map. We use two different noise levels $\sigma^{max}_n = 10\mu K$ and $\sigma^{max}_n = 30\mu K$.
The pixel space standard deviation of the noise field, $\sigma_n(\gamma)$, is shown in left of Fig.~\ref{fig:NoiseSD}
and a sample noise map, $ n(\gamma)$, is shown on the right of the same figure. 

We extract the BipoSH signal from the noisy map using SIToolBox. The analysis is done in ecliptic coordinate system.  Fig.~\ref{fig:isotropicmap} shows that our method can recover the  sky signals from the noisy skymap. As the noise variance has a quadrupolar structure we can see that the $A^{20}_{ll}$ and $A^{20}_{ll-2}$ BipoSH coefficients of the noisy map (Gray) are deviated from $0$ at high multipoles. However, the plots show that the analysis can recover all the BipoSH coefficients  even in case of high anisotropic noise. 
The BipoSH coefficients that we recover (Red)  using our algorithm match well with BipoSH coefficients of the intrinsic skymap (Brown). As the intrinsic CMB map is statistically isotropic we can see that the recovered BipoSH coefficients are consistent with $0$ within $1-2\sigma$. 
\begin{figure*}
\includegraphics[width=0.32\textwidth,trim =  0 90 0 80, clip]{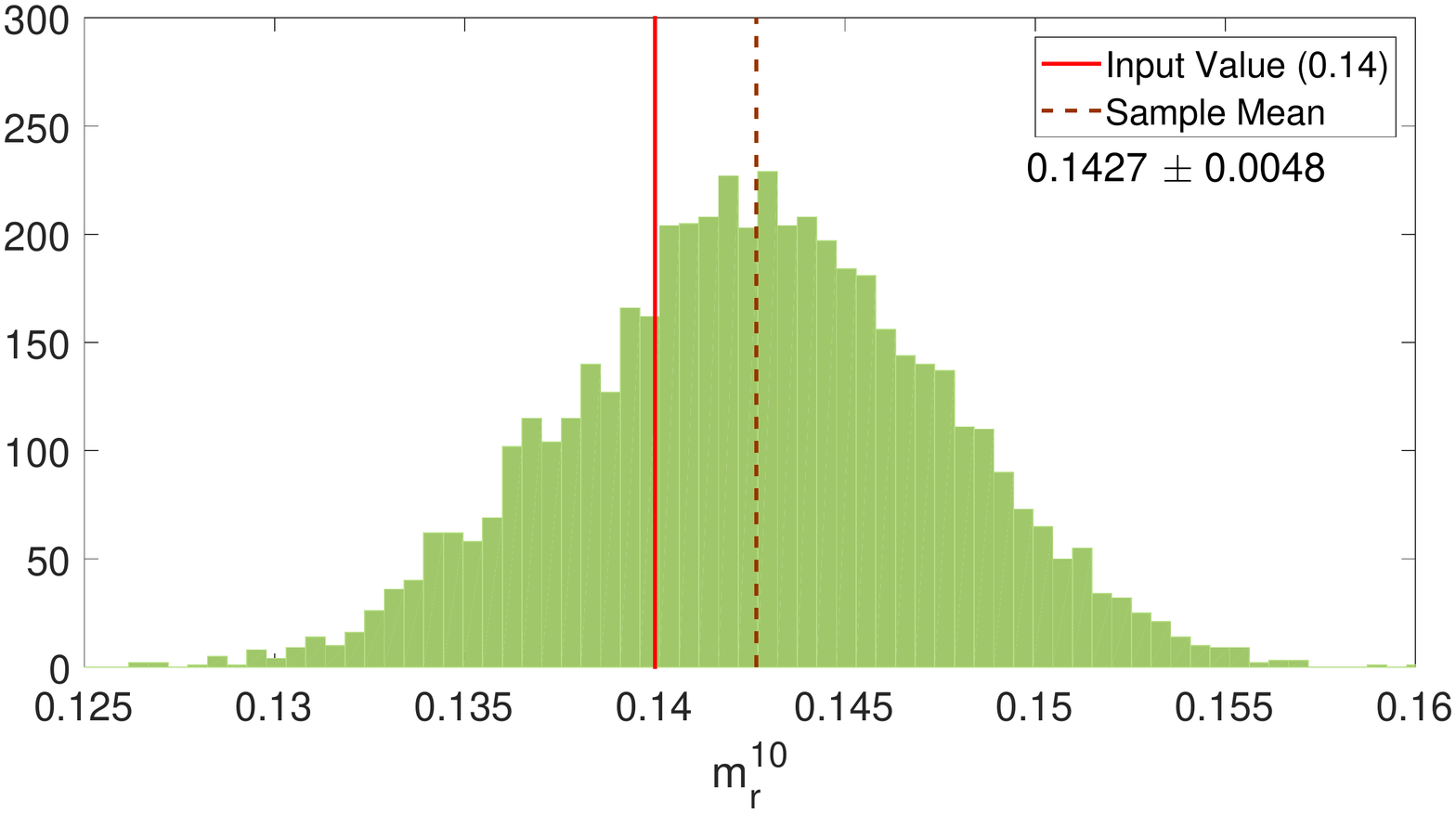}
\includegraphics[width=0.325\textwidth,trim = 0 90 0 80, clip]{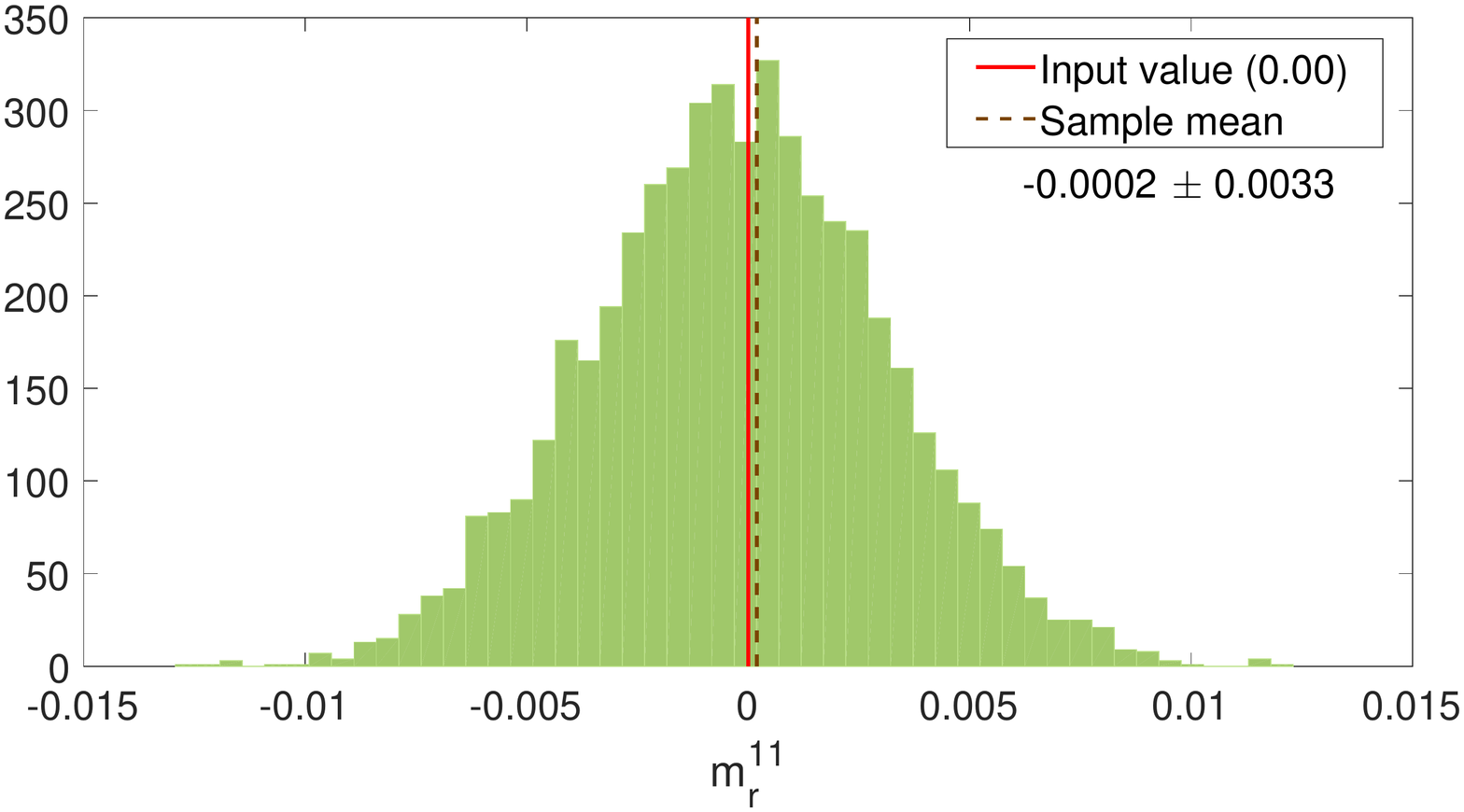}
\includegraphics[width=0.32\textwidth,trim = 0 90 0 80, clip]{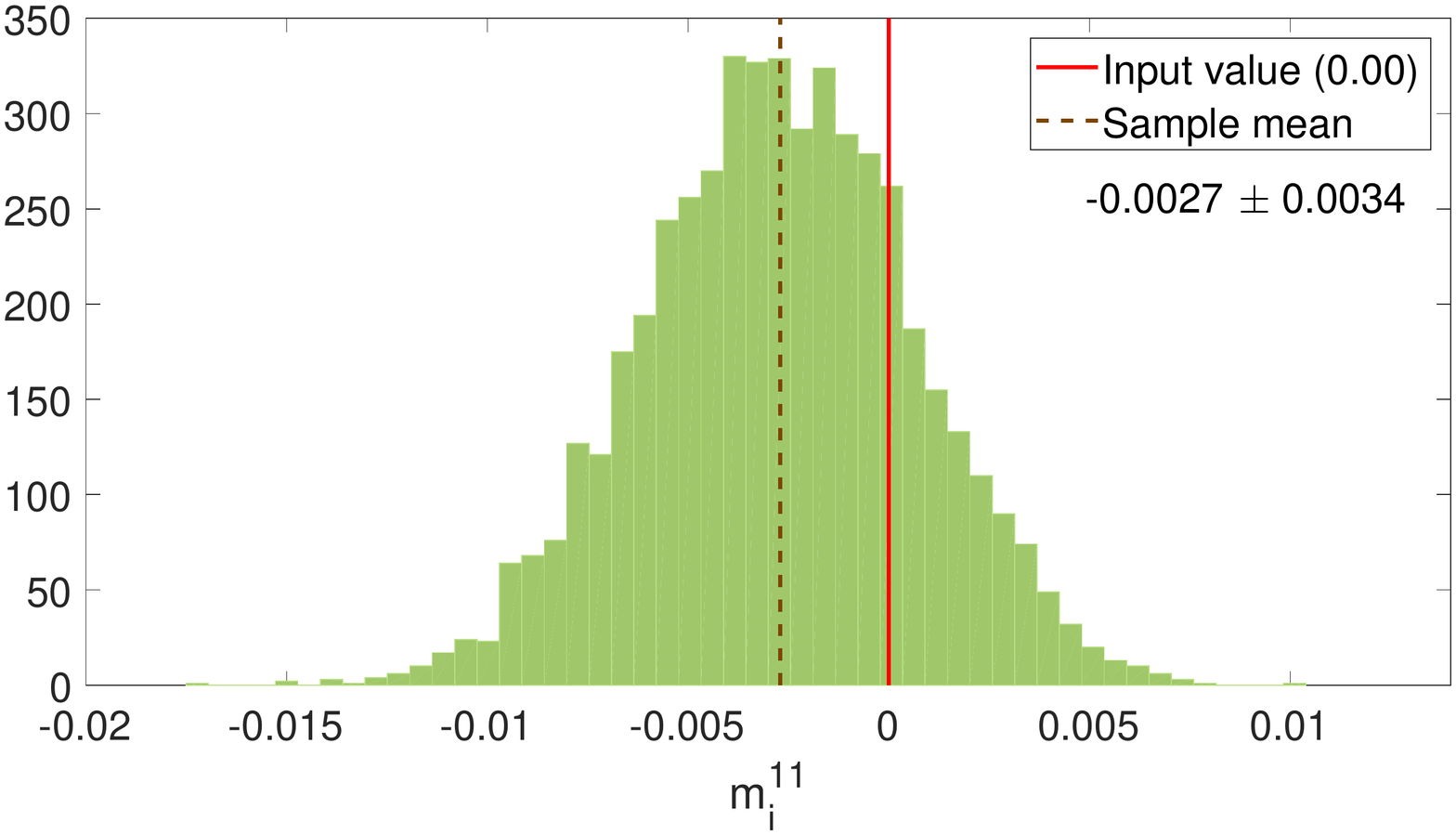}
\caption{The posterior of $m^{1M}$ is calculated from a realization, generated from input ${m^{10} = 14.0 \times 10^{-2}}$
and ${m^{11} = 0.0 + 0.0\;i}$, by directly sampling the likelihood. From the 
recovered posterior, we obtain ${m^{10} = 14.27\times10^{-2} \pm 4.80\times 10^{-3}}$ and ${ m^{11} = ( 2.0 \times 10^{-4} - 2.71\times10^{-3}i ) \pm ( 3.3 \times 10^{-3} + 3.4\times10^{-3}i )}$.}
\label{DipoleModulation}
\end{figure*}

\subsection{Anisotropic skymap and Anisotropic noise\label{ASAN}}

In a space based CMB experiment like WMAP or Planck, noncircular beam
coupled with the scan pattern can introduce SI violation in an otherwise SI
CMB sky~\citep{Das2016, Pant2016}. We take a SI skymap and then scan the map with WMAP W2~\footnote{\url{https://lambda.gsfc.nasa.gov/product/map/dr5/beam_maps_get.cfm}} band beam and scan-pattern and 
reconstructed the map from the time ordered data (TOD). The detail description of the map-making process can be found in~\citep{Das2016}. With this map we add similar anisotropic noise as discussed in the previous section. We estimate the BipoSH coefficients from the resultant skymap using SIToolBox.

In Fig.~\ref{fig:anisotropicmap}, we show the results for two different noise levels. The analysis is done in ecliptic coordinate system with HEALPix\footnote{\url{https://healpix.sourceforge.io/}} $N_{side}=512$ and $l_{max} = 1024$. For matching our results with the theory (Green) we generate $30$ W2 band beam convolved skymap for WMAP scan pattern from random SI realization generated with HEALPix. We calculate the BipoSH coefficients for all the 30 maps and take the average of those BipoSH coefficients as the theoretical value of the BipoSH. (An approximate semi-analytical approach for calculating the BipoSH coefficients with any beam shape and an arbitrary scan pattern can be found in \citep{Pant2016}). 
The plots show that our analysis can recover the isotropy violation signals in presence of high anisotropic noise. All the error-bars are matching with the theoretical results within 1-2$\sigma$. At high $l$ we can see slight difference between the theoretical BipoSH coefficients and the predicted value. This slight discrepancies are coming due to the properties of the particular realizations (Gray), which are slightly different from the average BipoSH coefficients (Green).

\begin{figure*}
\includegraphics[width=0.47\textwidth,trim = 0 0 0 0, clip]{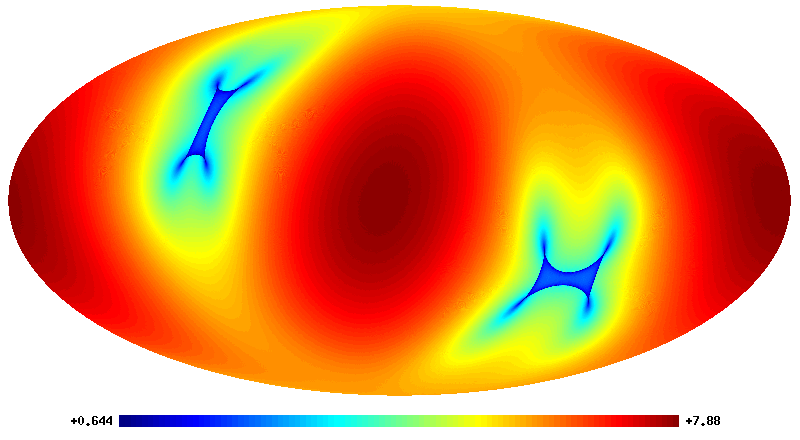}
\includegraphics[width=0.47\textwidth,trim = 0 0 0 0, clip]{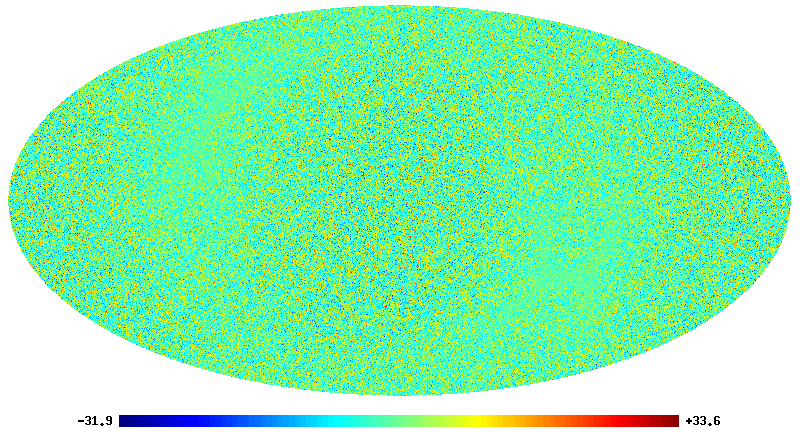}
\caption{Left: Noise standard deviation map ($\sigma_n(\gamma) $) used for the analysis. Right: A sample noise map,  generated from $\sigma_n(\gamma) $, i.e. used for the analysis. }
\label{Noise}
\end{figure*}

For generating the noise map, we use the hit count from a Planck like scan pattern 
As before we consider that the noise standard deviation map is inversely proportional to the square root of the number of times a pixel gets scanned (hit count), i.e. $\sigma_n(\gamma) = 8\mu K\times {\mathcal F_P}(\gamma)/{\mathcal F_P}_{max}(\gamma)$ where ${\mathcal F_P}(\gamma)=1/\sqrt{H_p(\gamma)}$. $H_P(\gamma)$  is the number of times a pixel along $\gamma$ direction get scanned in a Planck like scan-pattern. The noise standard deviation map is shown in the left of Fig.~\ref{Noise}. A sample noise map is shown in the right of the same figure.

\subsection{\label{maskedBiposh}Anisotropic skymap with anisotropic noise and masking}

Masking or the incomplete sky coverage 
provides a source of isotropy violation which is much bigger than other isotropy violation signals in the CMB sky. Therefore, it's important to remove the effect of masking in order 
to extract the SI violation signals from the background sky-map.
In this particular analysis, we apply our algorithm for the BipoSH calculation on masked sky.

Incorporating effect of masking in our analysis is straight forward. It can be done by setting the Noise variance of the masked pixels to infinity. 
However, we need to discard a significantly long chain as the `burn in' steps for recovering the underlying map from the masked region of the skymap. 
The analysis is carried out with a map from the set that is produced for the analysis in Sec.~\ref{ASAN}.
The map is in an ecliptic coordinate system,  $N_{side}=512$ and $l_{max}=1024$. 
In Fig.~\ref{fig:anisotropicmaskmap_maps} we show the map that is used for the analysis. The image on the top-left, shows the
 mask in Ecliptic coordinate system. Top-right image is showing the beam-convolved skymap before adding any noise and masking. Bottom-Left plot is for the masked noisy 
skymap ($\sigma_n^{max} = 30\mu K$).  We use similar noise standard deviation  map as that shown in Fig.~\ref{fig:NoiseSD}. This particular map is taken as the initial input value to the program. Our algorithm takes about $11,000$
samples  from a single chain for recovering the features in the masked region of the input skymap with each integration step about $\sim 3$ times longer 
than that is used in the unmasked analysis. Each of these integration steps involves about $60$, $S_{lml'm'}$ matrix inversion, and pixel 
to $a_{lm}$ and $a_{lm}$ to pixel space conversion. We had to discard all these starting samples as the `burn in' steps. One of recovered 
realization from post `burn in' step is shown in the Bottom-Right plot.

In Fig.~\ref{fig:anisotropicmaskmapspectra} we show the results from our analysis. We have taken $\sim 20,000$ post `burn in' samples for our analysis. 
The plots show that our analysis can recover the isotropy violation signals even in presence of masking. All the error-bars are matching with the theoretical results within 1-3$\sigma$. Therefore, even in presence of masking, where the BipoSH coefficients of the masked noisy skymap are significantly different from the intrinsic skymap, SIToolBox can recover the BipoSH coefficients reasonably well.

\begin{figure*}
\includegraphics[width=0.47\textwidth,trim = 0 0 0 0, clip]{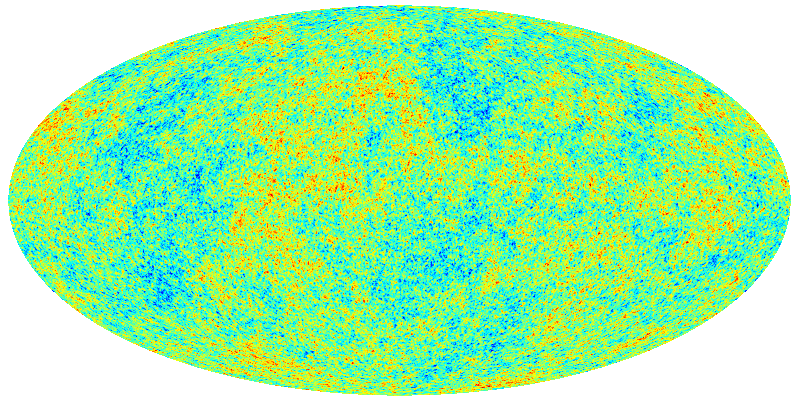}
\includegraphics[width=0.47\textwidth,trim = 0 0 0 0, clip]{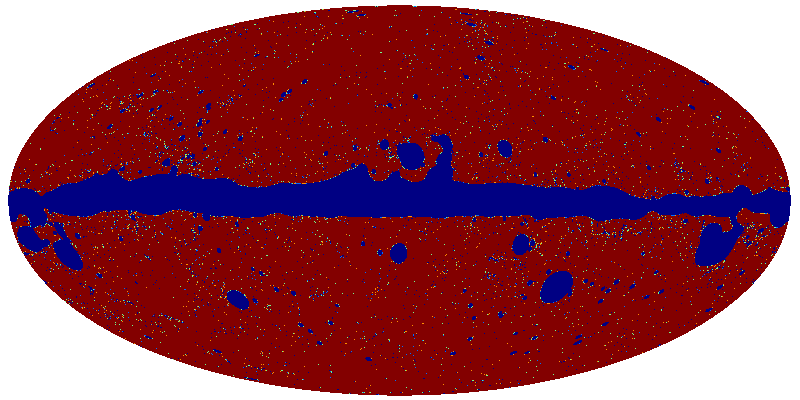}
\caption{Left: Original map with dipole modulation signal. Right: Mask map. The values in the mask map should be either $0$ or $1$. The fractional values are coming because the original mask map was in a higher resolution ($N_{side}=2048$). We downgrade it to a lower resolution ($N_{side} = 512$).}
\label{Mask}
\end{figure*}
\begin{figure*}
\includegraphics[width=0.47\textwidth,trim = 0 0 0 0, clip]{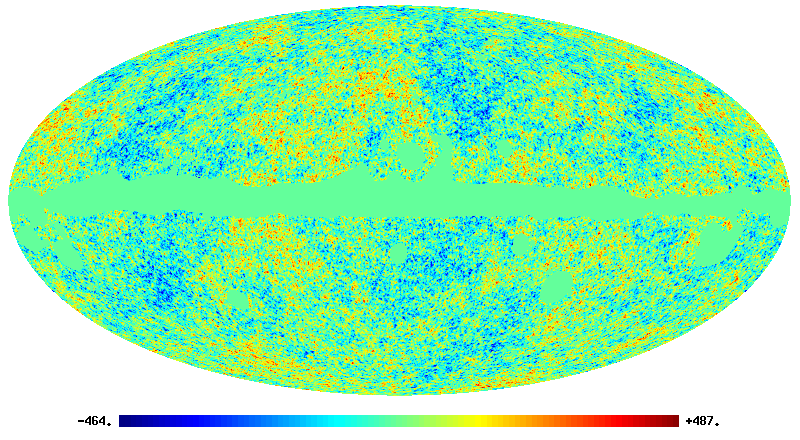}
\includegraphics[width=0.47\textwidth,trim = 0 0 0 0, clip]{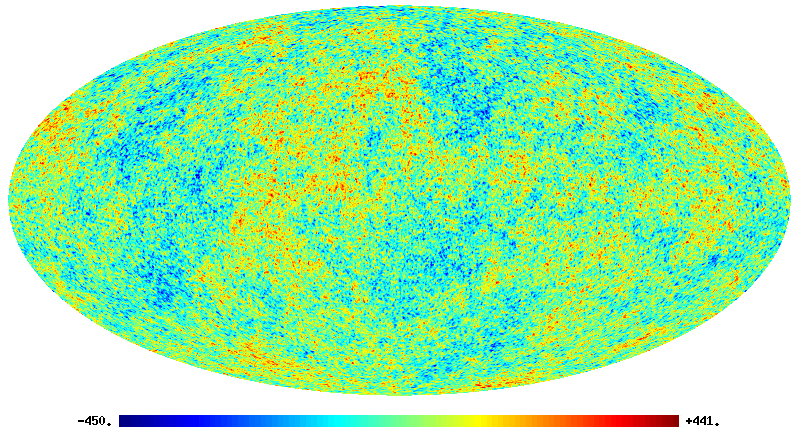}
\caption{Left: The masked noisy map used for recovering the dipole modulation signal. Right: The recovered map from one of the sample.}
\label{Recoveres}
\end{figure*}
\section{Dipole modulation in presence of anisotropic noise}

CMB  sky shows the hemispherical asymmetry which can be explained 
by modulating the low CMB multipoles with a dipole. There are different models of dipole modulation for explaining the hemispherical asymmetry. In some models the modulation amplitude is a function of the multipole numbers ($l$). However, in this analysis we consider a constant modulation amplitude for all the multipoles. 
We can produce a dipole modulated skymap by multiplying a SI skymap with a dipole as  
$T_{\text{dm}}(\gamma)=T_{\text{SI}}(1 +\sum_m m_{1m}Y_{1m}(\gamma))$. 
The resultant dipole modulated sky map will show SI violation in $A^{1m}_{ll'}$ BipoSH
coefficients~\citep{Mukherjee2013}.

Given a dipole modulated sky map, our goal is to calculate the dipole modulation amplitude and the posterior distribution, $P(m_{1m},a_{lm}|d_{lm})$.
The probability distribution for $A^{LM}_{ll'}$ will be same as Eq.(\ref{lastEquation}) except $A^{1M}_{ll-1}= m^{1m} f_s(l)$, $A^{1M}_{l-1l}= m^{1m} f_s(l)$ 
and for all other $L$ and $L\ne 0$, $A^{LM}_{ll'}= 0$. $L=0$ is essentially a scaled angular power spectrum, which should be nonzero. Here, $f_s(l)$ is called shape factor for dipole modulation and can be calculated analytically. For dipole modulation, the shape factor is given by

\begin{equation}
f_s(l) = \frac{\sqrt{(2l+1)(2l+3)}}{\sqrt{12\pi}}\Bigg( {\mathcal C}_l + {\mathcal C}_{l+1} \Bigg)C^{10}_{l0l+10} \,,
\end{equation}
\noindent where ${\mathcal C}_l$ is the CMB angular power spectrum and $C^{10}_{l0l+10}$ is the Clebsch-Gordan coefficient.

If we define a momentum corresponding to $m^{1m}$, i.e. $p_{m^{1m}}$, the equation of motion for $m^{1m}$ will be 

{\small
\begin{align}
\dot{p}_{m^{1m}} &= \frac{\partial \ln P(m_{1m},a_{lm}|d_{lm})}{\partial m^{1m}} = \frac{\partial \ln P(m_{1m},a_{lm}|d_{lm})}{\partial m^{1m}}  \nonumber\\
&= 2 \sum_{l}\frac{\partial  A^{1M}_{ll-1}}{\partial m^{1m}}\frac{\partial \ln P(m_{1m},a_{lm}|d_{lm})}{\partial  A^{1M}_{ll-1}} = 2\sum_{l}f_{s}( l )\dot{p}_{A^{1M}_{ll-1}}\;,
\end{align}
}

\noindent where $\dot{p}_{A^{1M}_{ll-1}}$ is given by Eq.(\ref{eq:Eq8}). For calculating the $\dot{m}^{1m} = \frac{p_{m^{1m}}}{m_{{m}^{1m}}}$, we need the mass matrix for ${m}^{1m}$ i.e.  $m_{{m}^{1m}}$. For our calculation we take the mass to be $m_{m^{1m}} = -\sum_l \frac{f_s(l)f_s(l+1)}{C_l C_{l+1}}$. The negative sign is important because in Hajian-Souradeep  format the $f_s(l)f_s(l+1)$ will always be negative. This mass matrix ensures the stability of the method.

\begin{figure*}
\includegraphics[width=0.32\textwidth,trim =  20 200 20 210, clip]{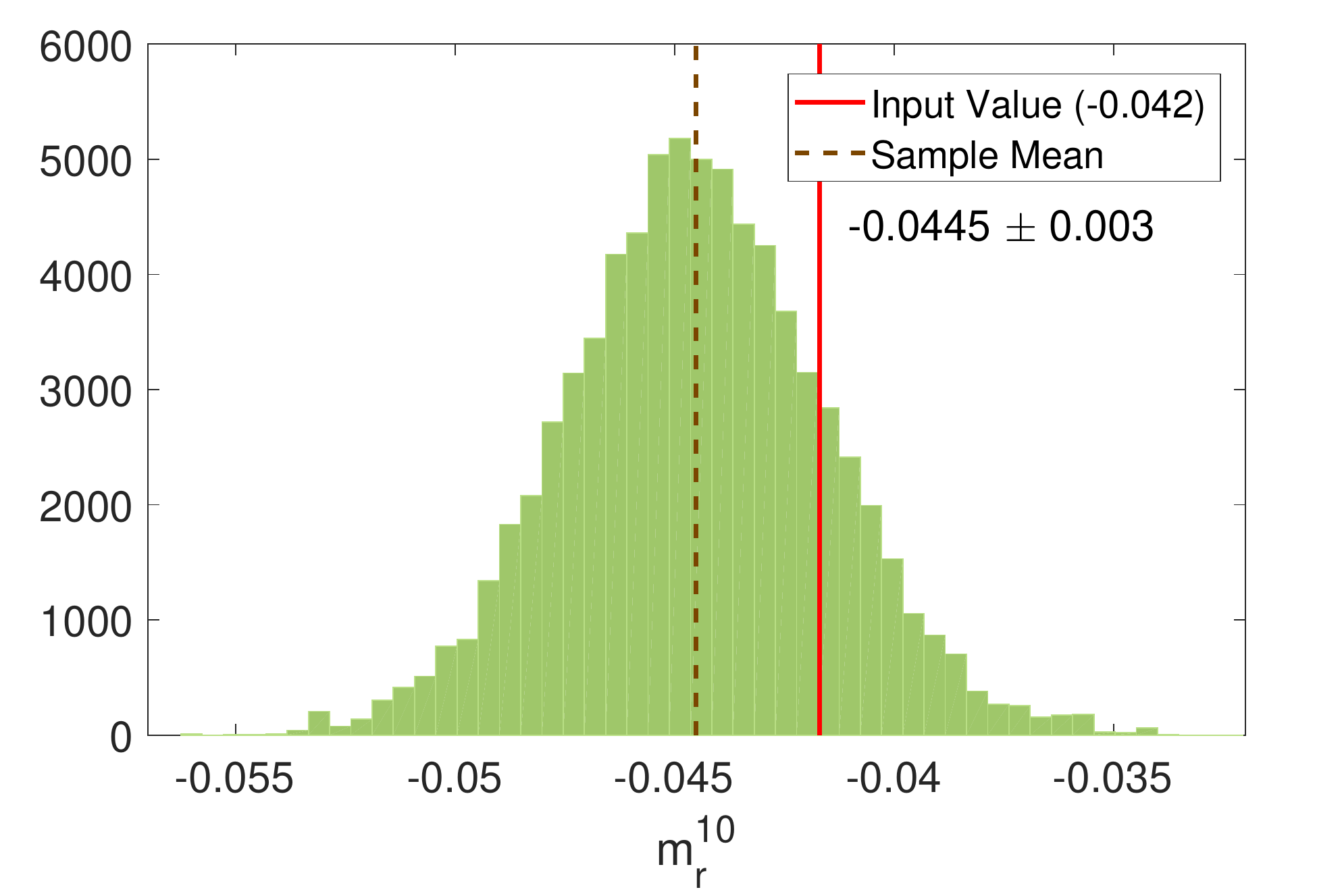}
\includegraphics[width=0.32\textwidth,trim =  20 200 20 210, clip]{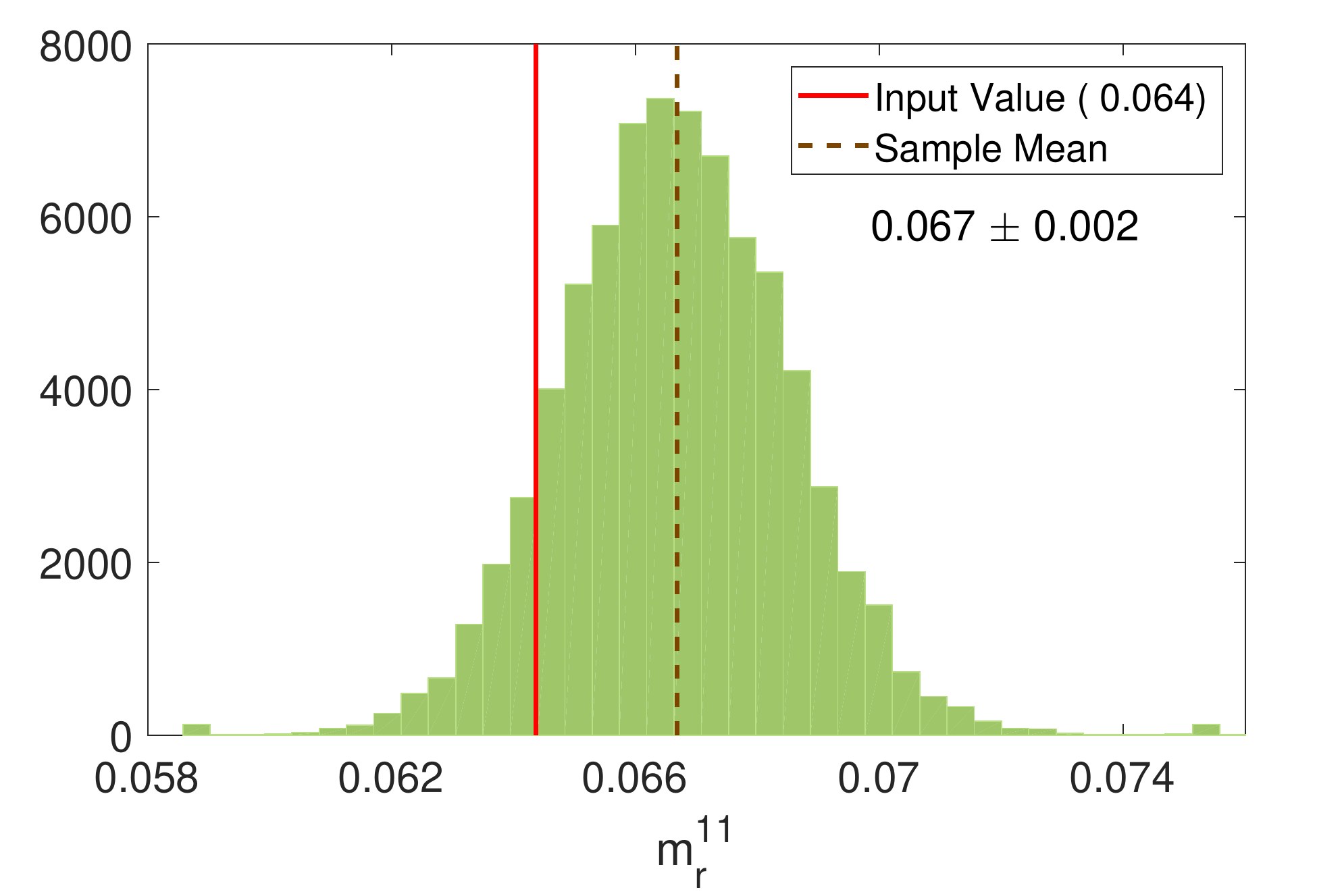}
\includegraphics[width=0.32\textwidth,trim =  20 200 20 210, clip]{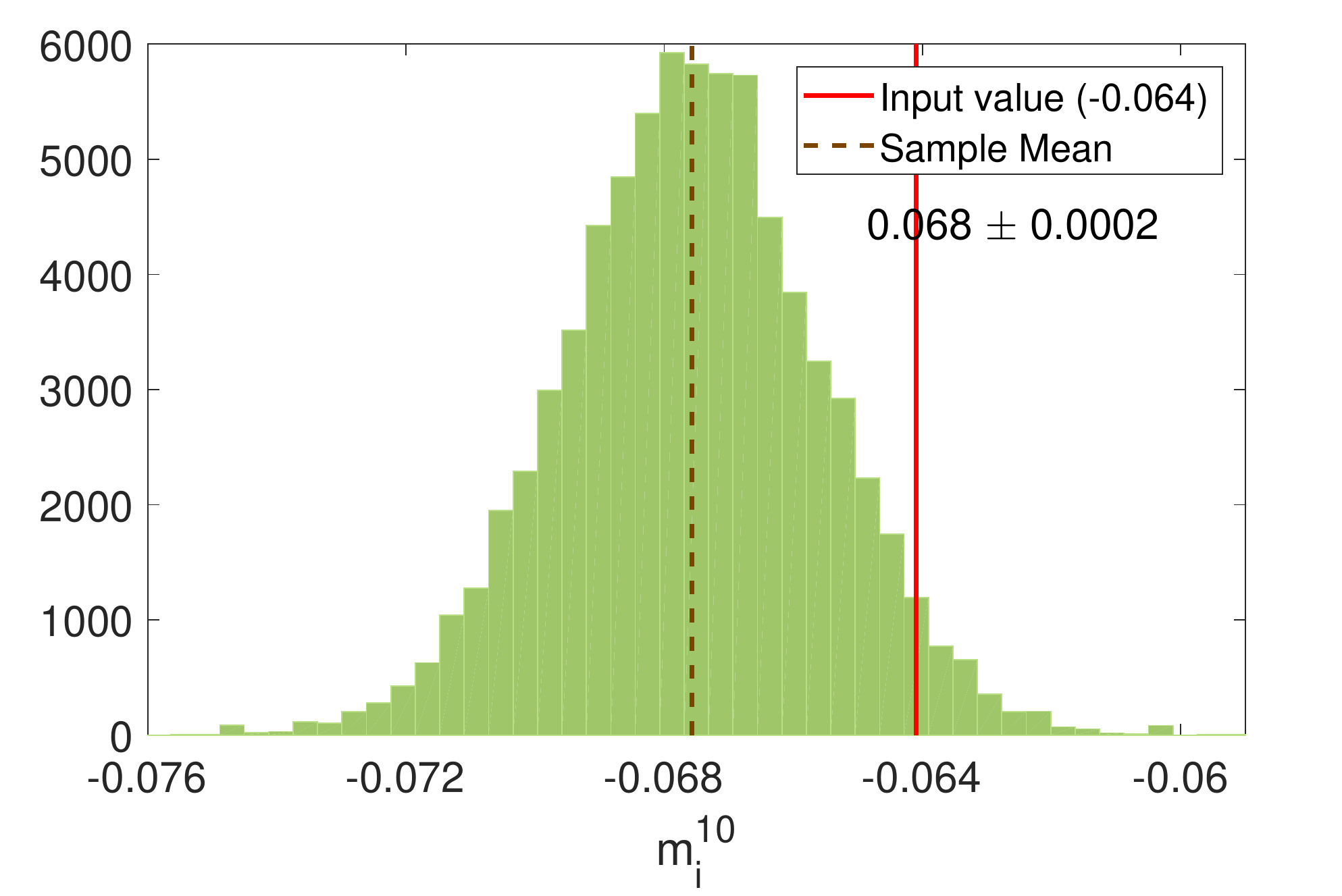}
\caption{The posterior of $m^{1M}$ is calculated from a realization, generated from input ${m^{10} = -0.42 }$
and ${ m^{11} = 0.0644 - 0.0641\;i}$, in presence of anisotropic noise and masking. From the 
recovered posterior, we obtain ${ m^{10} = -4.45\times10^{-2} \pm 3.00\times 10^{-3}}$ and ${m^{11} = ( 6.67 \times 10^{-2} - 6.76\times10^{-2}i ) \pm ( 1.93\times 10^{-3} + 2.0\times10^{-3}i )}$.}
\label{ShabbirDipoleModulation}
\end{figure*}

For our analysis, we modulate a SI realization produced using HEALPix, $N_{side}=512$ from a Planck like power spectrum,
 with a dipole sky map of $0.14\times Y_{10}(\gamma)$. This creates a nSI sky map with $A^{1m}_{ll-1} = m^{1m} f_s(l)$, 
where $m^{10} = 14.0 \times 10^{-2}$ and $m^{11} = 0.0 + 0.0\;i$. $l_{max} = 1024$ is used for the analysis.

We take $ \sigma_n(\gamma) = 5\mu K\times(1+Y^{10}(\gamma))$ as the noise standard deviation in the pixel space. 
We consider a dipolar noise matrix instead of a WMAP like profile as used before, because the earlier noise profile has a quadrupolar structure and will have almost no effect or a very little effect on the dipole modulation. So, we use a noise profile that 
can affect the result significantly. 

Fig.~\ref{DipoleModulation},  shows that SIToolBox can recover the input signal even in presence of high anisotropic noise. The recovered values from our analysis are $m^{10} = 14.27\times10^{-2} \pm 5.00\times 10^{-3}$ and
$m^{11} = ( 1.3 \times 10^{-4} - 3.3\times10^{-3}i ) \pm ( 3.5 \times 10^{-3} + 3.5\times10^{-3}i )$. 
All the recovered values are within $1\sigma$ of the input values.

\subsection{\label{dipolemodulationmasked}Calculating dipole modulation in presence of masking}

For this analysis we use a dipole, similar to the dipole measured by Planck 2015~\citep{Ade2016}. Dipole amplitude is $|\alpha| = 0.066$ and the modulation direction is 
$(\theta,\phi) =(225^{\circ}, -18^{\circ}$) in galactic coordinate system.  To represent the coefficients in terms of the spherical harmonics we can write $\vec{\alpha} = \sum_{-1}^{1}m_{1m}Y_{1m}(\theta,\phi)$.
Putting the expressions for $Y_{lm}(\theta,\phi)$ we obtain

\begin{equation}
\begin{aligned}
m_{10} = \sqrt{\frac{4\pi}{3}}\alpha\cos(\theta), \;\;\;\;\;\;  \frac{m^i_{11}}{m^{r}_{11}} = - \tan(\phi), \;\;\;\;\;\; \\
 \alpha^2 = \frac{3}{4\pi}\Bigg( m_{10}^2 + 2{m^r_{11}}^2 + 2{m^i_{11}}^2\Bigg)\,.
\end{aligned}
\end{equation}

Replacing the values for $\alpha$, and $(\theta,\phi)$ we get $m_{10} = -0.0417$, $m_{11}^r = .0644$ and $m_{11}^{i} = -0.0641$. For generating the modulated skymap 
we use CoNIGS (a software package developed by \citep{Mukherjee2013} for producing Gaussian nSI realizations from a given shape factor). For masking we use the SMICA\footnote{\url{https://irsa.ipac.caltech.edu/data/Planck/release_2/ancillary-data/}}~\citep{Cardoso2008,Ade2014a} mask from Planck analysis. In Fig.~\ref{Mask}, we show the original map (left) and the mask (right), used for this analysis. The original mask map was in $N_{side} = 2048$. We have downgraded it to $N_{side}=512$. The fractional values in the mask originates from the downgrading process. In our analysis we set the mask values to $1$ if it is larger than $0.5$, otherwise set it to $0$. 

 In Fig.~\ref{Recoveres} we show the noisy skymap after masking (left). 
One of realization from the the recovered samples  is shown in the right of the same figure.
To reduce the burn in steps, we use noisy un-masked skymap as the starting value for the HMC Sampling. 

In Fig.~\ref{ShabbirDipoleModulation} we show the dipole modulation parameters recovered from the masked sky. The plots show that the recovered value of 
$m^{10} = -4.45 \times 10^{-2} \pm 3.0\times 10^{-3}$ and $m^{11} = ( 6.7 \times 10^{-2} - 6.8\times10^{-2}i ) \pm ( 1.93 \times 10^{-3} + 2.0\times10^{-2}i )$.
All the values are within $1-2\sigma$ of the input signal showing that the algorithm for estimating dipole modulation works for partial sky coverage.

\begin{figure*}
\includegraphics[width=0.32\textwidth,trim = 80 90 80 100, clip]{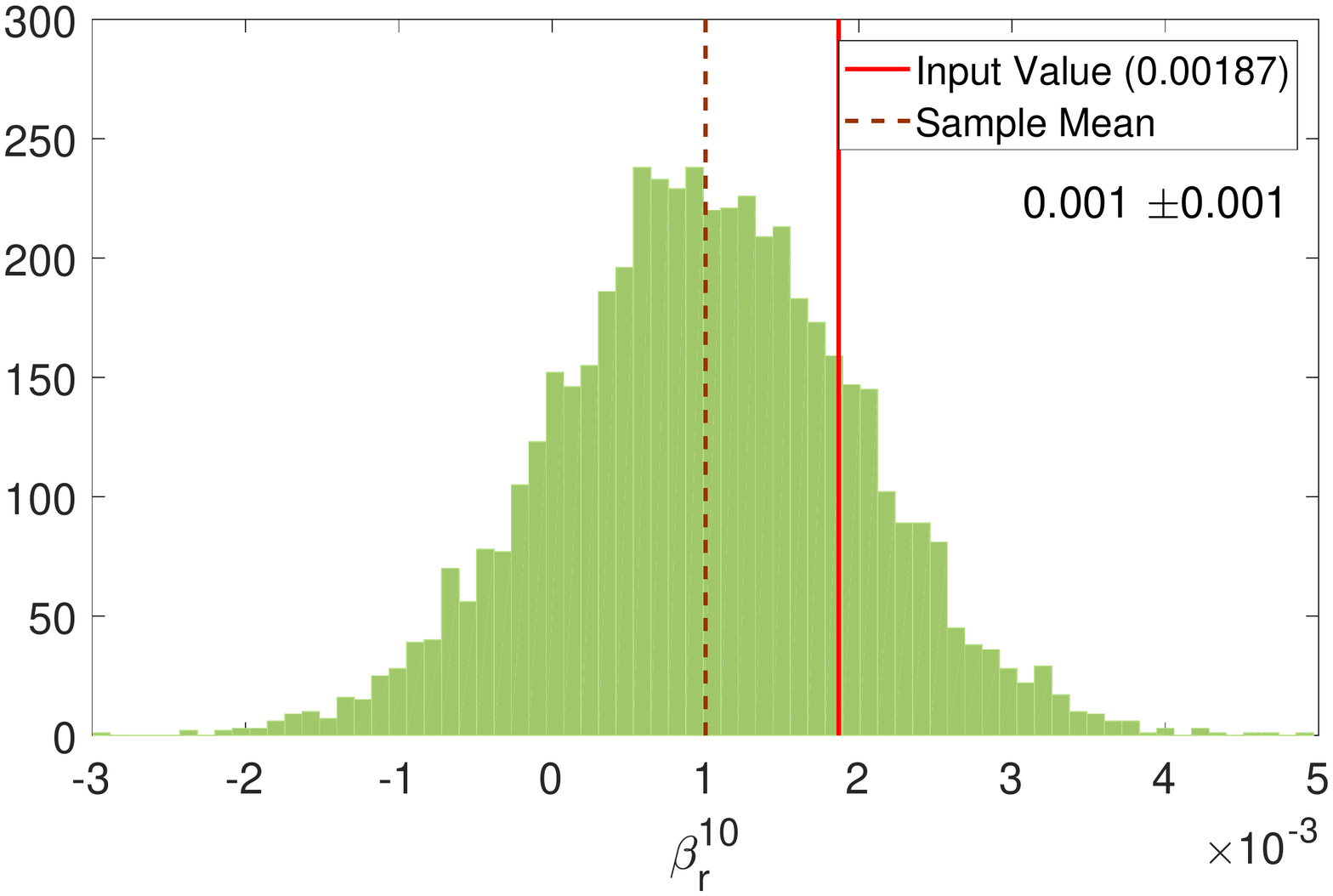}
\includegraphics[width=0.32\textwidth,trim = 80 90 80 100, clip]{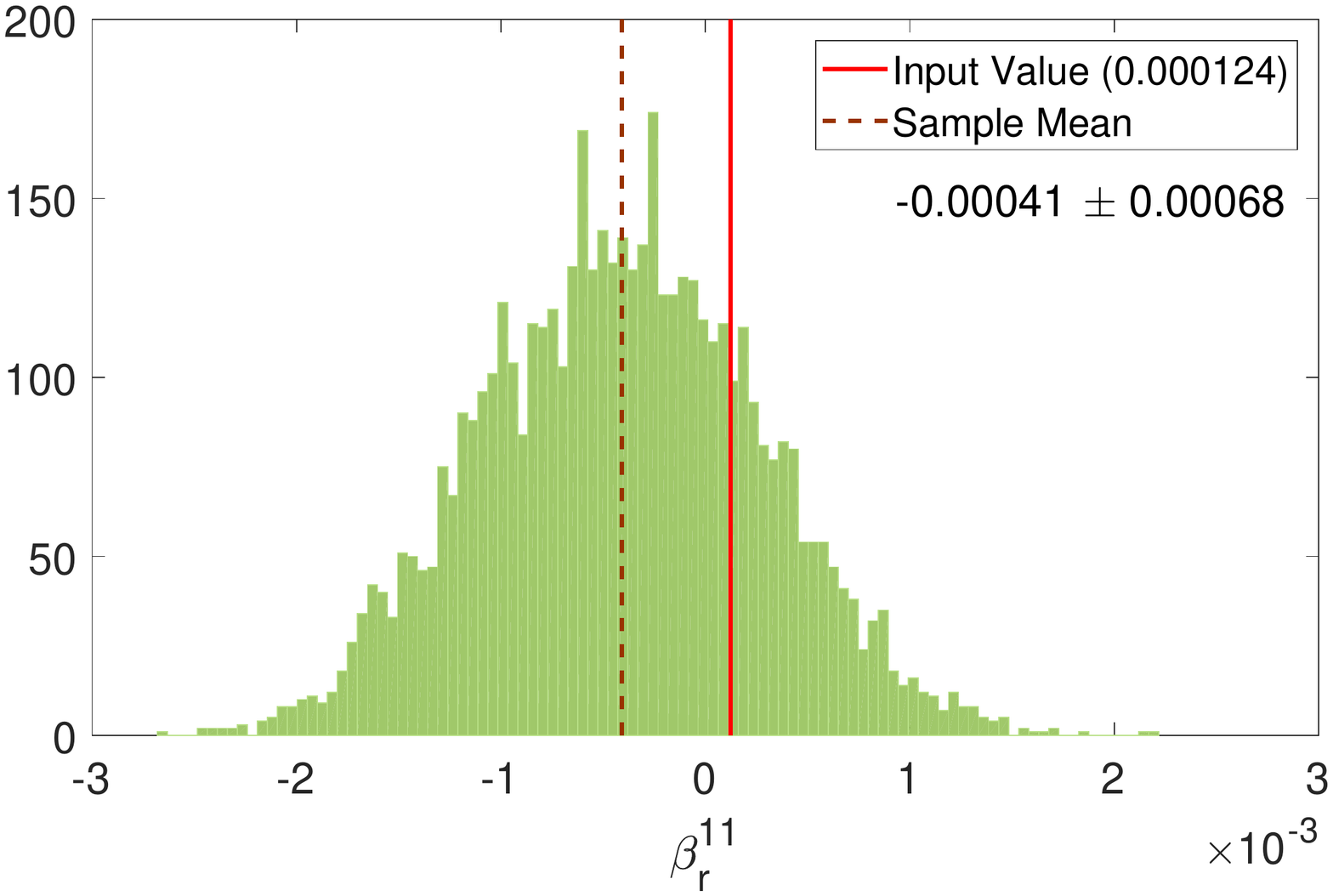}
\includegraphics[width=0.32\textwidth,trim = 80 90 80 100, clip]{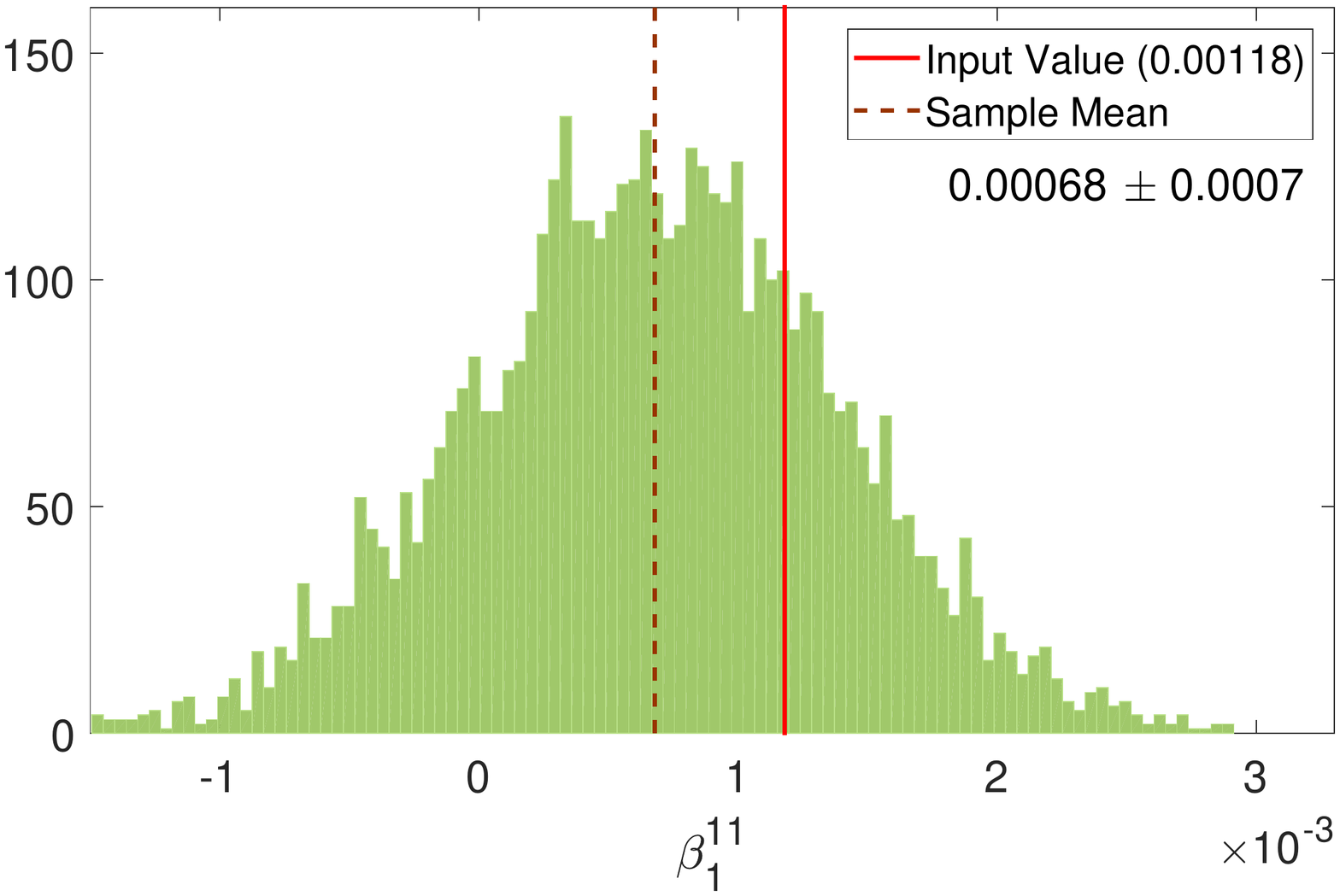}
\caption{The posterior of $\beta^{1M}$ is calculated from a realization, generated from input ${\beta^{10}=1.87\times10^{-3}}$
and ${\beta^{11}=-1.24\times10^{-4}+1.18\times10^{-3}i}$, by directly sampling  the likelihood. From the 
recovered posterior, we get ${\beta^{10} = 1.00\times10^{-3} \pm 9.76\times 10^{-4}}$ and 
${\beta^{11} = (- 4.1 \times 10^{-4} + 2.78\times10^{-4}i ) \pm ( 6.8 \times 10^{-4} + 7.1\times10^{-4}i )}$.}
\label{DopplerBoost}
\end{figure*}

\section{Doppler boost parameters in presence of anisotropic noise}

Another well known source of the isotropy violation in the CMB
signal is the Doppler boost, caused by the motion of our galaxy with
respect to the CMB rest frame. A well known feature of this Doppler boost is visible on the
CMB monopole and gives rise to a high CMB dipole which we need to
subtract from the CMB signal during data analysis. However, the Doppler
boost does not only change the signal of the CMB monopole, but also modifies 
the higher multipoles of CMB which leads to the
SI violation.

A robust data analysis technique should be able to measure the Doppler boost from SI violation
signal, which can then be compared with the Doppler boost detected from
the CMB dipole and check if there is a mismatch. 
A detailed discussion on  
Doppler boost can be found in \citep{Mukherjee2014,Mukherjee2013}. The Doppler Boost
 leads to aberration in the direction of incoming photons and also modulation of the Stokes parameter and combined 
effect is given by the shape factor
{\small
\begin{equation}
f_s(l)=\frac{\sqrt{(2l+1)(2l+3)}}{\sqrt{12\pi}}\Bigg[(1+b_\nu){\mathcal C}_l^{TT} - (l+2-b_\nu){\mathcal C}_{l+1}^{TT}\Bigg]C^{10}_{l0l+10}\,,
\end{equation}
}
\noindent where $b_\nu$ is a frequency dependent quantity and is given by 
\begin{equation}
b_\nu = \frac{\nu}{\nu_0}\coth\Bigg(\frac{\nu}{\nu_0}\Bigg) - 1\,. 
\end{equation}


The algorithm for extracting Doppler boost parameters is exactly same as dipole modulation, except the shape factor for the 
Doppler boost is different from the dipole modulation. Also, the Doppler boost signal is stronger at high multipole.

We construct a nSI CMB sky map using 
CoNIGS~\citep{Mukherjee2013}, where we inject
a Doppler boost signal with $\beta^{10}=-1.87\times10^{-3}$ and $\beta^{11}=-1.24\times10^{-4}+1.18\times10^{-3}i$. We use $N_{side} = 512$ and $l_{max} = 1024$ for this analysis.  

We add an anisotropic noise with standard deviation $ \sigma_n(\gamma) = 5\mu K\times(1+Y^{10}(\gamma))$ 
and run SIToolBox to estimate the values of $\beta^{1M}$ parameters. The recovered values from our algorithm are $\beta^{10} = 1.00\times10^{-3} \pm 9.76\times 10^{-4}$ and $\beta^{11} = (- 4.1 \times 10^{-4} + 2.78\times10^{-4}i ) \pm ( 6.8 \times 10^{-4} + 7.1\times10^{-4}i )$.
Plots are shown in Fig.~\ref{DopplerBoost}. We can see that the recovered signal matches with the injected
values with $1 - 2\sigma$.


\section{\label{discussion} Discussion and Conclusion}
In this paper we extend our previous work of estimating the underlying co-variance structure on a sphere, for anisotropic noise and partial sky coverage. This makes the algorithm more suitable for application on real data. We use HMC method for estimating the BipoSH coefficients from the CMB sky-map in presence of different noise profiles and masking.
SIToolBox is able to successfully recover the full CMB BipoSH
signal up to $l_{max} = 1024$ with good accuracy. 
For the BipoSH 
calculations we use $\sigma_n^{max} = 10\mu K$ and 
$\sigma_n^{max} = 30\mu K$ for $N_{side}=512$  (pixel size $6.8 \, arcmin$), which are much higher than then noise
in any present CMB experiments like Planck where $\sigma_n^{max} \sim 4\mu K$\footnote{\url{https://crd.lbl.gov/departments/computational-science/c3/c3-research/cosmic-microwave-background/cmb-data-at-nersc/}}. 
In future experiments like CMB-S4 the fourcasted white noise level is about $1 \mu K - arcmin$.

We also carry out a direct Bayesian inference of the posterior distribution of the the Doppler boost parameter ($\beta$) and dipole modulation signals observed in CMB sky. Our algorithm can recover the injected signals effectively from the simulated anisotropic skymap. As the Doppler boost signal is stronger at higher multipoles, while analyzing a real skymap map, choosing $l_{max}\sim 2000$ or a higher $N_{side}$ can provide better results. Our algorithm is capable of doing the analysis up to any given $l_{max}$. However, in such cases the computation time will also increase as $\sim N_{side}^4$, making the process highly time consuming.
Also, in our analysis we expand the co-variance matrix in terms of Taylor series under the assumption that the covariance matrix is diagonally dominant. 
However, at very high $l$, if  the assumption of statistical isotropy brakes down due to lensing then the Taylor series expansion may 
require higher order terms to produce accurate results. 

We face another challenge while analyzing the masked maps. HMC should theoretically work for any initial value. The parameters should first converge towards the best-fit value and then sample the distribution around it. Convergence is very fast with constrained data set. The burn-in sample size is small and we can run multiple HMC chains independently in parallel and get a large number of samples. However, in presence of masking, the constrain on the higher multipoles in the masked region only comes from the unmasked part of the sky. Therefore, the convergence is slow.  If masked sky is taken as the initial value then it takes significantly large number of samples to recover the map of the masked region. We need to discard a significantly large number of initial samples as the burn-in step. Therefore running multiple chains to get large number of samples is not a convenient, as from each of the chains we need to discard large number of sample points as burn-in. Hence, the process is time consuming. It takes $\sim5,000$ CPU hours (single chain on 16 OpenMP cores) for simulating Sec~\ref{maskedBiposh}. On the other hand, our analysis show that if we start with an unmasked noisy sky as the input value, the process stabilizes much faster. This allows us to run multiple parallel chains with significantly less burn-in samples (see Sec~\ref{dipolemodulationmasked}). However, SIToolBox can recover the BipoSH coefficients with any starting value given a significantly long chain.  

Recently Planck releases CMB polarization results on isotropy~\citep{Collaboration2019}. Our algorithm is readily applicable to the CMB polarization maps for analyzing the isotropy violation. 
SIToolBox can, in principle, be used for understanding the covariance structure of any random field over a sphere and not restricted to the CMB application. Another emerging  field in astronomy is the HI intensity mapping. Several intensity mapping surveys, like Tianlai~\citep{Das2018,Chen2012}, HIRAX, FAST, GBT etc. are either mapping or planning to map the $21cm$ sky signal. SIToolBox can be used directly to analyze the data. Apart from that other areas of research, like research in geoscience, climate modeling etc, also use random field over a sphere and SIToolBox can help them with their research. 
 

\section*{Acknowledgement}
Work at UW-Madison and Fermilab is supported by NSF Award AST-1616554. This research is performed using the computer resources and assistance of the UW-Madison Center For High Throughput Computing (CHTC) in the Department of Computer Sciences. The CHTC is supported by UW-Madison, the Advanced Computing Initiative, the Wisconsin Alumni Research Foundation, the Wisconsin Institutes for Discovery, and the National Science Foundation, and is an active member of the Open Science Grid, which is supported by the National Science Foundation and the U.S. Department of Energy's Office of Science. Author wish to thank Shabbir Shaikh for many fruitful discussions and for his sincere help in testing algorithm with various inputs. Author thanks Shabbir Shaikh and Suvodip Mukherjee for helping in fixing a missing factor of 2, which was present in the previous paper~\citep{Das2015}.  Author also wish to thank Tarun Souradeep and Benjamin D. Wandelt for many useful discussions throughout the course of this project.  

\bibliographystyle{mnras}
\bibliography{SI_toolbox_ref}

\end{document}